\documentclass[usenatbib,12pt,letterpaper,notitlepage]{article}
\usepackage{times}
\usepackage[margin=1in]{geometry}
\usepackage{graphicx}
\usepackage{amssymb}
\usepackage{amsmath}
\usepackage{amsbsy}
\usepackage{epstopdf}
\usepackage{pdflscape}
\usepackage{setspace}
\usepackage{nicefrac}
\usepackage{multirow}
\usepackage{booktabs}
\usepackage{natbib}
\usepackage{enumerate}
\usepackage{makeidx}  
\usepackage[titletoc]{appendix}
\usepackage{fancyhdr}
\usepackage{longtable}
\usepackage[latin1]{inputenc}
\usepackage{geometry}
\usepackage{graphicx}
\usepackage{epstopdf}
\usepackage{amsmath}
\usepackage{amsfonts}
\usepackage{amssymb}
\usepackage[parfill]{parskip}	
\usepackage{color}
\usepackage{bm}
\usepackage{float}
\usepackage{setspace}		
\onehalfspace
\usepackage{hyperref}
\usepackage{subfigure}

\usepackage[font=small,labelfont=bf]{caption} 

\onehalfspace
\newcommand{\ub}[1]{_{\bar{\text{#1}}}}			
\newcommand{\unb}[1]{_{\text{#1}}}				
\newcommand{\tP}[0]{\text{P}}					

\allowdisplaybreaks

\begin{document}

\title{\begin{large}
Estimating the Size and Distribution of\\Networked Populations with Snowball Sampling\thanks{
This work was supported through a Natural Sciences and Engineering Research Council Postgraduate Scholarship D and a Discovery Grant. The authors wish to thank Laura Cowen, Charmaine Dean, Ove Frank, Maren Hansen, Chris Henry, Kim Huynh, Richard Lockhart, and Carl Schwarz for their helpful comments. The authors would also like to thank John Potterat and Steve Muth for making the Colorado Springs data available.}
\end{large}
}
\author{\begin{large}Kyle Vincent\end{large}\footnote{Independent Researcher, Ottawa, Ontario, CANADA,  \textit{email}: kyle.shane.vincent@gmail.com}
\begin{large}and Steve Thompson\end{large}\footnote{Department of Statistics and Actuarial Science,
Simon Fraser University, 8888 University Drive,
Burnaby, British Columbia, CANADA, V5A 1S6,
\textit{email}: thompson@sfu.ca}\newline
}

\date{\today}

\maketitle

\begin{abstract}
\begin{small}
\noindent
A new strategy is introduced for estimating population size and networked population characteristics. Sample selection is based on a multi-wave snowball sampling design. A generalized stochastic block model is posited for the population's network graph. Inference is based on a Bayesian data augmentation procedure. Applications are provided to an empirical and simulated populations. The results demonstrate that statistically efficient estimates of the size and distribution of the population can be achieved.
\end{small}
\end{abstract}

\thispagestyle{empty}

\pagenumbering{arabic}

\clearpage

\section{Introduction}

Hard-to-reach populations are typically not covered by a sampling frame, thereby making recruitment a challenge for the investigator. As there is usually a contact pattern between the members of such populations, link-tracing sampling designs can be used to exploit the underlying network to find members to sample for a study. Hence, there has been a growing interest in the use of link-tracing sampling designs to facilitate recruitment for studies based on these populations. A recent surge in the literature relates to inference procedures oriented around such designs for estimating attributes of networked populations. We contribute to this area by developing a new strategy based on a stratified setup, multi-wave snowball sampling design, and model-based approach to inference to estimate the full population size. The strategy also has the ability to estimate model parameters that govern strata assignments and presence of links between individuals.

In our approach we base recruitment on the snowball sampling design due to its 1) popularity from a theoretical standpoint \citep{Frank2009, Frank2011} and 2) its practicality for recruitment in an empirical setting \citep{Browne2005, Petersen2011}. In our theoretical setup we generalize the stochastic block model for random graphs \citep{Nowicki2001}. The model is a direct extension of the Bernoulli graph model in that it can account for stratum assignments and allows for links to occur independently between individuals conditional on such assignments. In our setup we opt to use this model as it 1) requires a minimal amount of observations on individuals selected for the sample, in the form of categorical covariate information for stratum assignments and not necessarily continuous covariate information, which in turn can reduce the amount of resources required for a study, 2) can serve as a good omnibus model for social networks since it is a relatively simple model that has a considerable amount of application to hard-to-reach populations \citep{Thompson2000, Thompson2003}, and 3) results in a tractable likelihood-based estimation procedure for the population size when a multi-wave snowball sampling design is used.

Direct maximum likelihood estimation based on likelihoods arising from network models like those based on exponential random graph models (ERGMs) and/or induced through network samples can be analytically or computationally burdensome. One approach is to use a Bayesian data augmentation routine as this can provide a more viable alternative to inference \citep{Kwanisai2004, Koskinen2013}. In this study we present in full detail an extended strategy allowing for estimation of the population size that is made possible by utilizing the labels-ignored likelihood. Similar to the ERGM-based approaches of \cite{Pattison2013} and \cite{Rolls2013}, our approach has the ability to make inference on graph parameters when the population size is unknown, as well as to estimate the population size. Furthermore, the strategy has the advantage in that links between nodes in the final wave are not required to be observed.

Contemporary approaches to network-based estimates include the following. The network scale-up method, recently empirically evaluated by \cite{Salganik2011}, is based on a random sample of the general population and indirect observations about personal networks that intersect with the target population. The method possesses several advantages over other approaches. For example, it does not require respondents to disclose their membership, or the membership of those in their network, if in the target population. Also, it is logistically simple and inexpensive to implement in practice. However, the network scale-up method requires strong assumptions. For example, the networks of those in the general population are assumed to be representative of the population. In contrast, our approach can permit for, and therefore exploit for inferential purposes, differing patterns of nominations within and between strata.

Bayes-aided approaches based on respondent driven sampling (RDS) designs are presented in \cite{Handcock2014} and \cite{Handcock2015}, as well as \cite{Crawford2018}. In the former two they use a successive sampling approximation with selection probability proportional to degree to model the RDS process, but otherwise not its network characteristics. They then include their approximating design in the Markov chain Monte Carlo (MCMC) inference calculations. In the latter they use a network approach based on the assumption of a Bernoulli graph model. They first infer on the links within the sample (since the RDS design does not make such observations) and then use those in a Bayes-based estimation procedure with a likelihood based on the binomial distribution applied to the approximated number of links from each sampled individual to those outside the sample. RDS typically requires many waves of recruitment for inferential procedures to be applied. In contrast, our approach does not necessarily require more than one wave of sampling for the corresponding inference procedure to be applied.

\cite{Vincent2017} develop a design-based approach, which bases preliminary estimation on the \cite{Frank1994} estimators for a one-sample study and mark-recapture estimators \citep{Rivest2001} for a multi-sample study. This approach exploits a sufficiency result and Rao-Blackwellization to incorporate individuals selected through a link-tracing design to improve on the preliminary estimates. The design is based on adding one individual linked to the current sample at each phase of selection, but not to complete multi-wave snowball sampling designs.

The new strategy presented in this paper is based on selecting an initial Bernoulli sample and then tracing links out to add several waves of individuals to the sample. Through positing a stochastic block model based on strata assignments, estimation of the population size, model parameters that govern the probability of assignment of individuals to each strata, and model parameters that govern the probability of links arising within and between pairs of individuals within/across each stratum/strata, is made possible. To evaluate the new method we conduct a variety of simulations using a network data set as an empirical population, as well as network data sets generated from the stochastic block model. Relative to existing approaches, we find that, in general, the Bayes data augmentation estimators for the population size often perform as well or better than other estimators. When the stochastic block model fits the network well, the gains in improvement over these estimators are substantial. The use of more than one wave has the potential to significantly reduce the bias and/or variance. In cases where the stochastic block model does not fit the network well, we find that there is some robustness to the new strategy as it can provide less biased estimators. Finally, in cases where sampling commences with selecting the more central individuals of the population, we find that the benefit of using the new approach when based on multi-waves gives estimates with reasonable mean squared error.

\section{Graph Model Setup and Sampling Design}

In this section we first generalize the stochastic block model. We then reintroduce the multi-wave snowball sampling design.

\subsection{The stochastic block model}

\cite{Thompson2000} explore the use of the stochastic two-block model when sampling is based on a snowball sampling design. We follow their setup and generalize the model to work over as many blocks as desired. Hereafter, we refer to blocks as strata.

First, we posit that all units in the population $U=(1,2,...,N)$ are independently assigned to one of $G$ strata, which are latent for all individuals unless sampled, with a probability corresponding to $\underline{\lambda}=(\lambda_1,\lambda_2,...,\lambda_G)$ where $\lambda_j$ is the probability an assignment is made to stratum $j$ for $j=1,2,...,G$.  We define $C_i$ to be the stratum to which individual $i$ belongs. Second, for simplicity, in our study we assume that all links are reciprocated; we define $Y$ to be the symmetric adjacency matrix of the population where for all $i,j=1,2,...,N, Y_{ij}=Y_{ji}=1$ if a link is present between units $i$ and $j$, and 0 otherwise. We posit that conditional on the population vector of stratum memberships $\underline{C}=(C_1,C_2,...,C_N)$, links occur independently between all pairs of units in the population; for any $i,j=1,2,...,N$, $\text{if}\ i\neq j$ then $P(Y_{ij}=1|\underline{C})=P(Y_{ij}=1|C_i,C_j)=\beta_{C_i,C_j}$, and if $i=j$ then $P(Y_{ii}=1)=0$. It shall be understood that for all $k,\ell=1,2,...,G$, $\beta_{k,\ell}=\beta_{\ell,k}$.

Stratum assignments should be carried out according to factors of importance, that is, those which explain/predict the pattern of social links between individuals of the population. Social links between individuals in hard-to-reach populations, such as those comprised of drug-users, are typically based on the mutual sharing of drugs, drug-using equipment, and/or sexual relationships. Hence, covariate information that can explain the pattern of such links would typically come in the form of drug-using habits and/or a combination of demographics like age, gender, and race of the pairs of individuals. The presence of such non-directed social links in the population will give rise to the symmetric adjacency matrix that indicates the presence of links between individuals.

\subsection{The Multi-Wave Snowball Sampling Design}

Hard-to-reach populations are typically not covered by a sampling frame. Consequently, the investigator will not necessarily have complete control over the sample selection procedure. Therefore, one approach is to model the initial sample selection procedure as if it arises from a Bernoulli sampling design; see \cite{Frank1994} for further details. Hence, we assume the design commences with the selection of an initial sample/wave $S_0$ via such a design. All links are traced out to the corresponding nominations of those individuals. Those nominations outside the initial sample comprise the first wave. Sampling continues in this pattern until $W$ waves are reached, where all links are traced from each wave. Those individuals added at wave $w$ are denoted as $S_w$ for $w=0,1,...,W$.

Let $S=\cup_{w=0}^WS_w$ denote the final sample, and $S\setminus S_W=\cup_{w=0}^{W-1}S_w$ denote the final sample minus those individuals selected for the final wave. The data observed from the sample is $d_{Obs}=\{S_0,S_1,...,S_W,\underline{C}_{S_0},\underline{C}_{S_1},...,\underline{C}_{S_W},Y_{S\setminus S_W,S},Y_{S\setminus S_W,\bar{S}}\}=\{S,\underline{C}_{S},Y_{S\setminus S_W,S},
Y_{S\setminus S_W,\bar{S}}\}$ where $\bar{S}=U\setminus S$ is the set of members not selected for the final sample; $\underline{C}_S$ is the vector of the observed stratum memberships of the sampled members;  $Y_{S\setminus S_W,S}$ refers to the recorded observations on the presence and absence of links between the first $W-1$ waves and the final sample; and $Y_{S\setminus S_W,\bar{S}}\equiv 0$ is understood to be the absence of links between the individuals selected for the first $W-1$ waves and those individuals not selected for the final sample (for which there is an unknown number).

\section{The Likelihoods}

In this section we first present the full likelihood, which is based on a full graph realization. We then evaluate the observed likelihood and labels-ignored likelihood based on the sample data.

\subsection{The Full Likelihood}

Under the generalized stochastic block model, the likelihood function for the population parameters based on an entire graph realization is
\begin{align}
&L(\underline{\lambda},\underline{\beta}|\underline{C},Y)=\prod\limits_{k=1}^G \lambda_k^{N_k} \prod\limits_{k=1}^G\beta_{k,k}^{M_{k,k}}\prod\limits_{k=1}^G(1-\beta_{k,k})^{{N_k \choose 2}-M_{k,k}}
\prod\limits_{\underset{k<\ell}{k,\ell=1:}}^G\beta_{k,\ell}^{M_{k,\ell}}\prod\limits_{\underset{k<\ell}{k,\ell=1:}}^G
(1-\beta_{k,\ell})^{N_k N_\ell-M_{k,\ell}}
\label{full likelihood block}
\end{align}
\noindent where $N_k$ is the size of stratum $k$ and $M_{k,\ell}$ is the number of links within/between the members of strata $k$ and $\ell$, for $k,\ell =1,2,...,G$. The first component of the likelihood corresponds to the assignment of the stratum memberships to the individuals, the second and third components correspond to the assignment of links within each stratum, and the fourth and fifth components correspond to the assignment of links between strata.

\subsection{The Observed Likelihood}

\cite{Thompson1996} and \cite{Thompson2000} present the mathematical details for obtaining the observed likelihood when a model-based approach to inference is used and selection is based on a link-tracing design. Here we generalize those results for unknown population size to obtain the observed likelihood.  To facilitate inference we condition on the size of the initial Bernoulli sample/wave so that inference proceeds as if the sample is obtained with a simple random sampling design. Conveniently setting $n_w=|S_w|$, $n=n_0+n_1+...+n_W$, and recall that $S\setminus S_W=\cup_{w=0}^{W-1}S_w$, we express the \textit{observed likelihood} as
\begin{align}
&L_{Obs}(N,\underline{\lambda},\underline{\beta}|d_{Obs},|S_0|) \notag\\
&=p(S|N,|S_0|,Y_{S\setminus S_W,U})\times\sum\limits_{C_{\bar{S}},Y_{S_{W}\cup\bar{S},S_{W}\cup\bar{S}}} f(\underline{C},Y|N,\underline{\lambda},\underline{\beta})\notag\\
&=p(S_0|N,|S_0|) \times \sum\limits_{C_{\bar{S}},Y_{S_{W}\cup\bar{S},S_{W}\cup\bar{S}}}
f(\underline{C},Y|N,\underline{\lambda},\underline{\beta})\notag\\
&= \frac{1}{{N \choose n_0}}\prod\limits_{i\epsilon S\setminus S_W}\lambda_{C_i}\times \prod\limits_{\underset{i<j}{i,j\epsilon S\setminus S_W:}} \bigg(  \left( \beta_{C_i,C_j} \right)  ^{Y_{ij}} \cdot \left(   1 - \beta_{C_i,C_j}  \right)^{(1-Y_{ij})}\bigg) \times\notag\\
&\prod\limits_{j\epsilon S_W}\bigg[\lambda_{C_j}\prod\limits_{i\epsilon S\setminus S_W} \bigg(  \left( \beta_{C_i,C_j}  \right)  ^{Y_{ij}}\cdot\left(   1 - \beta_{C_i,C_j}   \right)^{(1-Y_{ij})}  \bigg)\bigg] \times\bigg[\sum\limits_{k=1}^G\bigg(\lambda_k \prod\limits_{i\epsilon S\setminus S_W} \left(1- \beta_{C_i,k}\right)\bigg)\bigg]^{N-n}, \notag\\
\label{observedlikelihood}
\end{align}
\noindent where the last component of the likelihood is equivalent to $\prod\limits_{j\epsilon\bar{S}}\bigg[\sum\limits_{k=1}^G\bigg(\lambda_k\prod\limits_{i\epsilon S\setminus S_W} \left( 1-\beta_{C_i,k}\right)\bigg)\bigg]$.

The first component corresponds to the sampling design, the second corresponds to the observations made on the stratum memberships of the individuals selected for the first $W-1$ waves of the sample, the third corresponds to the observations made on links within the first $W-1$ waves of the sample, the fourth corresponds to the observations made on the stratum memberships of the last wave and then links between the first $W-1$ waves and the last wave, and the final component corresponds to both the unobserved stratum memberships of the individuals outside the final sample and the observed absence of links between these members and the first $W-1$ waves.

\subsection{The Labels-Ignored Likelihood}

The observed likelihood based on $d_{Obs}$ will not provide meaningful inference for the population size; if all parameters are held constant then the likelihood is a monotonically decreasing function of $N$ when $N\geq n$. Instead, we can ignore unit labels to derive a more suitable likelihood for the population size and model parameters;  see \cite{Royall1968}, \cite{Scott1973}, and \cite{Cassel1977} for discussions on how such an approach can lead to meaningful inference in some sampling contexts, and \cite{Williams2002} on how it leads to likelihood-based estimation in mark-recapture contexts.

We shall let $d_{Ign}=\{\underline{C}_{S_0},\underline{C}_{S_1},,...,\underline{C}_{S_W},Y_{S\setminus S_W,S},Y_{S\setminus S_W,\bar{S}}\}$ be the corresponding observed data, where unit labels are ignored and stratum memberships and the adjacency matrix are known only up to permutations of the original observations. Recall that $n=n_0+n_1+...+n_W$. The population can then be partitioned into $W+2$ sets corresponding to $S_0,S_1,,...,S_W$ and $\bar{S}$, each of size $n_0,n_1,...,n_W$, and $N-n$, respectively, in ${N \choose {n_0,n_1,...,n_W,N-n}}$ ways. Hence, the resulting \textit{labels-ignored likelihood} is
\begin{align}
&L_{Ign}(N,\underline{\lambda},\underline{\beta}|d_{Ign},|S_0|)\notag\\
&={{N-n_0} \choose n_1, ..., n_W, N-n} \times \prod\limits_{i=1}^{n-n_W}\lambda_{C_i} \times \prod\limits_{\underset{i<j}{i,j=1,2,...,n-n_W}:}\bigg(\beta_{C_i,C_j}^{Y_{ij}}(1-\beta_{C_i,C_j})^{(1-Y_{ij})}\bigg)\times\notag\\
&\bigg[\prod\limits_{j=n-n_W+1}^{n}\bigg(\lambda_{C_j}\prod\limits_{i=1}^{n-n_W}\bigg(\beta_{C_i,C_j}^{Y_{ij}}(1-\beta_{C_i,C_j})^{(1-Y_{ij})}\bigg)\bigg)\bigg]\times
\bigg[\sum\limits_{k=1}^G\bigg(\lambda_k\prod\limits_{i=1}^{n-n_W}(1-\beta_{C_i,k})\bigg)\bigg]^{N-n},
\label{blockignorelikelihood}
\end{align}
\noindent where we make use of the labels $1,2,...,n$, solely for presentation purposes (that is, only the structure of the observed subset of the graph is retained in this likelihood).

In contrast to the observed likelihood, which depends on unit labels, the labels-ignored likelihood is now one that is more suitable for estimating $N$. By definition of the stochastic block model, links between $S\setminus S_W$ and $U\setminus(S\setminus S_W)$ occur independently given the corresponding stratum memberships of those in $S\setminus S_W$ and $U\setminus(S\setminus S_W)$. These Bernoulli-type outcomes can be regarded as independent and, by symmetry, identically distributed. Hence, embedded within the labels-ignored likelihood is a binomial type of experiment where a ``success" occurs if a unit outside $S\setminus S_W$ is linked to at least one unit in $S_{W-1}$ and a ``failure" occurs otherwise; consider the initial and final terms in the likelihood.

\section{Data Augmentation}

When the population size is known, it is straightforward to analytically sum/integrate the corresponding likelihood over the unobserved part of the network; see \cite{Thompson2003} for further details. However, when the population size is unknown and the number of strata is large, maximum likelihood- or Bayes-based estimation of the population size and model parameters can be theoretically or computationally cumbersome; consider the complicated form of the likelihood presented in Expression \ref{blockignorelikelihood}. A Bayesian data augmentation procedure \citep{Tanner1987} based on a Gibbs sampling approach is therefore used for estimation since it is straightforward to evaluate the theoretical distributions of the model parameters and missing data alike, given the observed data. The procedure is based on iteratively sampling from the posterior conditional distributions, as detailed in this section. We resort to using this imputation procedure, in contrast to one which is based on summing/integrating over the unobserved part of the network, because 1) when the number of strata increases, the number of combinations to sum over can be analytically burdensome, and 2) when the population size is unknown the likelihood presented in Expression \ref{blockignorelikelihood} is a function of the population size, and hence the number of combinations grows exponentially while part 1) must be repeated for each value of the unknown population size.

\subsection{Imputation}

\subsubsection{Population Size}

We define the prior distribution on $N$ to be a power-law like distribution where $\pi(N)\propto N^{-a}$, $a=0,1,2,...$. When $a=0$ the prior distribution is uniform and improper, whereas when $a > 0$ it is a monotonically decreasing function that is improper. In all cases the prior distribution is unbounded. The resulting posterior distribution of $N$, conditional on $d_{Ign}$ and the most recently sampled model parameters $\underline{\lambda}$ and $\underline{\beta}$, is
\begin{align}
P(N|d_{Ign},\underline{\lambda},\underline{\beta})&=\frac{P(d_{Ign}|N,\underline{\lambda},\underline{\beta})\pi(N)}{P(d_{Ign}|\underline{\lambda},\underline{\beta})} \propto\frac{{{N-n_0} \choose n_1, ..., n_W,N-n}(1-p)^{N-n}\frac{1}{N^{a}}}{\sum\limits_{N'\geq n}\bigg({{N'-n_0} \choose n_1, ..., n_W, N'-n}(1-p)^{N'-n}\frac{1}{N'^{a}}\bigg)}\notag\\
&\propto {{N-n_0} \choose n_1,...,n_W,N-n}(1-p)^{N-n}\frac{1}{N^{a}}
\label{postNblock}
\end{align}
\noindent where $1-p=\sum\limits_{k=1}^G\bigg(\lambda_{k}\prod\limits_{i=1}^{n-n_W}(1-\beta_{C_i,k})\bigg)$. \noindent Notice the resemblance of Expression \eqref{postNblock} to the Binomial distribution, as remarked upon after presenting the labels-ignored likelihood in the previous section. Further, it is interesting to note that the distribution of $n_W$ given the composition of the first $W-1$ waves bears some resemblance to this result as $n_W|\underline{C}_{S\setminus S_W} \sim$ Binomial$(N-n+n_W,p')$ where $p'=\sum\limits_{k=1}^G\bigg(\lambda_{k}\prod\limits_{\ell=1}^{G}(1-(1-\beta_{k,\ell})^{n'_{\ell}})\bigg)$ and $n'_{\ell}$ is the number of individuals obtained for $S\setminus S_W$ from stratum $\ell$.

\subsubsection{Stratum Memberships}

The augmentation procedure now makes use of the labels $1,2,...,N$ solely for imputation purposes and continues with the observed graph data labeled as $d_{Obs'} = \{\text{S}, \underline{C}\unb{S},Y_{S\setminus S_W,U}\}$, where $U$ is a hypothetical population of size equal to the imputed value based on the distribution presented in Expression \eqref{postNblock}. We show in the appendix that for any $i \in \bar{\text{S}}$ and for any stratum $k = 1, 2, ... \, G$,
\begin{align}
&P(C_i = k | \text{S}, \underline{C}_{S\setminus S_W}, Y_{S\setminus S_W, i})=\frac{\lambda_k\prod\limits_{j=1}^{n-n_W}(1-\beta_{C_j,k})}
{\sum\limits_{\ell=1}^G\bigg(\lambda_l\prod\limits_{j=1}^{n-n_W}(1-\beta_{C_j,\ell})\bigg)} \text{ .}
\label{AA2}
\end{align}
\noindent Values of the missing stratum memberships $\underline{C}_{\bar{S}}$ can then be assigned according to the distribution outlined in Expression \eqref{AA2}.

\subsubsection{Links}

After $\underline{C}\ub{S}$ is imputed based on the distribution presented in Expression \eqref{AA2}, the graph data is updated from $d_{Obs'}$ to $d_1$ where $d_1=\{S,\underline{C},Y_{S\setminus S_W,U}\}$ and $\underline{C}$ represents a hypothetical full graph realization of stratum memberships. Now, for any $i,j\ \epsilon\ S_W\cup\bar{S}$ where $i\neq j$, links arise independently via $P(Y_{ij}=1|d_1)=\beta_{C_i,C_j}$ and can be assigned using this result. This results in a hypothetical full graph realization of $d=\{\underline{C},Y\}$.

\subsection{Posterior Distributions of Model Parameters}
\label{Posterior Distributions of Model Parameters}

With the use of independent prior distributions on the model parameters we use the factorization theorem to sample the model parameters under the hypothetical full graph realization $d=\{\underline{C},Y\}$; see the likelihood in Expression \eqref{full likelihood block}. In our study we place independent conjugate Dirichlet and Beta priors on $\underline{\lambda}$ and $\underline{\beta}$, respectively, as follows: $\pi (\underline{\lambda}) \propto \prod\limits_{k=1}^G \lambda_k^{\alpha_k-1}$ and $\pi(\underline{\beta}) \propto \prod\limits_{\underset{k\leq \ell}{k,\ell=1:}}^G\bigg(\beta_{k,\ell}^{\gamma_1-1}(1-\beta_{k,\ell})^{\gamma_2-1}\bigg).$

We take the prior distributions to be noninformative by setting $\alpha_k=1$ for $k=1,2,...,G$ and $\gamma_j=1$ for $j=1,2$. The resulting posterior distribution of $\underline{\lambda}$ is then $\pi(\underline{\lambda}|d) \sim \text{Dirichlet}(N_1+1,...,N_G+1)$. Similarly, the resulting posterior distribution of $\beta_{k,\ell}\ \text{for}\ k,\ell=1,2,...,G, k\neq \ell$ is $\pi(\beta_{k,\ell}|d)\sim \text{Beta} (M_{k,\ell}+1,N_kN_\ell-M_{k,\ell}+1)$, and for $k=1,2,...,G$ is $\pi(\beta_{k,k}|d) \sim \text{Beta} (M_{k,k}+1,{N_k \choose 2}-M_{k,k}+1)$.

The Dirichlet and Beta priors respectively correspond to one hypothetical individual/observation from each stratum, and one hypothetical link observation within/between strata. These priors are chosen to stabilize the inference procedure since a Bernoulli initial sample may not select individuals from particular strata, which could easily happen when stratum and/or sample sizes are small. Increasing the Dirichlet prior parameter values will have the effect of pulling the strata sizes closer together. For moderately sized samples, larger prior parameter values are likely to be required to have a discernible effect on the Bayes population size estimates.

\section{Empirical Study}

In this section the results from an application of the new strategy to an empirical population are presented. The purpose of the empirical study is to gauge the performance of the new strategy in an empirical setting relative to other strategies, as well as to determine its sensitivity to departures from the network model. Results are presented in the supplementary materials for a set of studies conducted on simulated populations.

\begin{figure}[H]
	\centering
\vspace{-1mm}
\centering
		
		\includegraphics[scale=0.35]{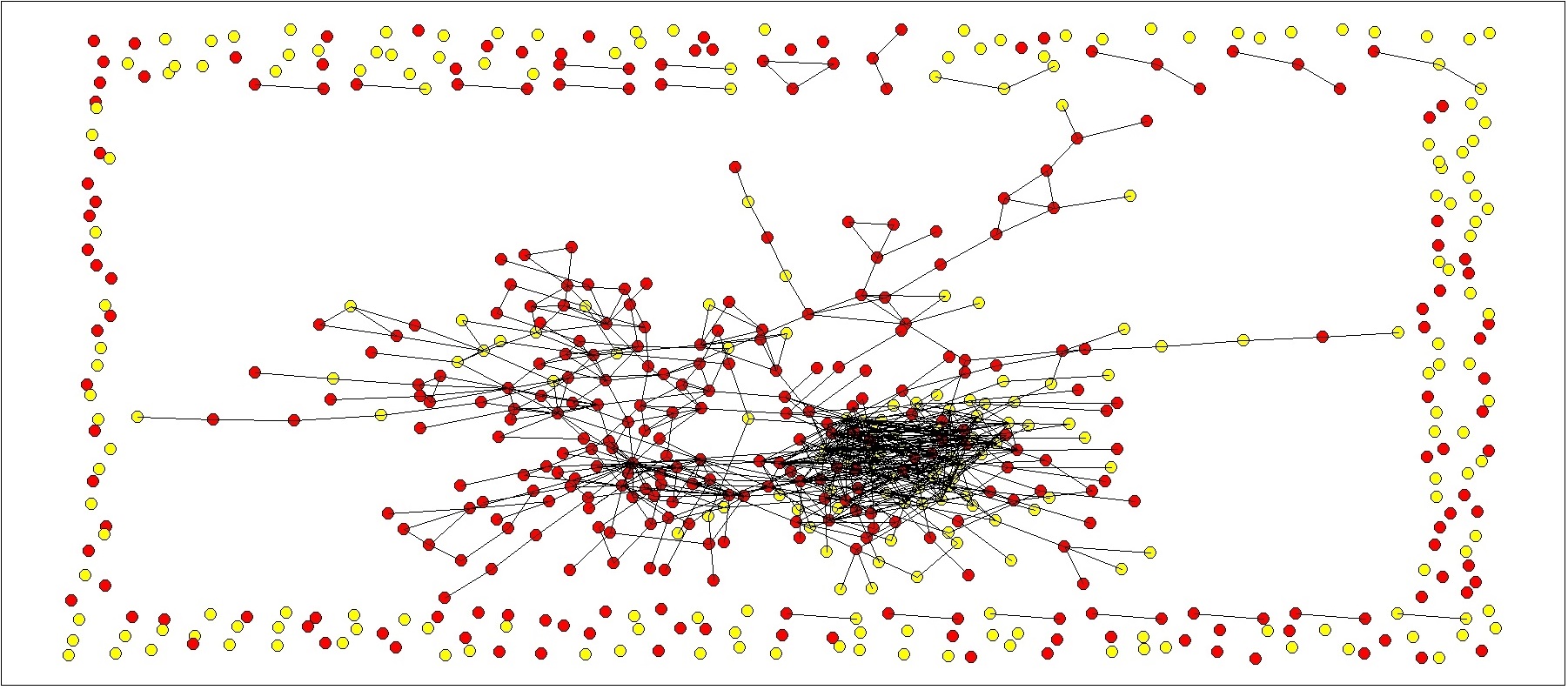}
\vspace{-3mm}
\caption{Empirical Population: Population of drug-users in the Colorado Springs area. The size of the population is 595 and the proportion of injection drug-users is 0.575.}
\label{CSpop}
\end{figure}

The empirical study is based on the P90 Colorado Springs study of 595 individuals at risk for HIV/AIDS \citep{Darrow1999, Potterat1993, Rothenberg1995}. Figure \ref{CSpop} illustrates the population network. The dark-coloured nodes represent the injection drug-users, and light-coloured nodes represent the non-injection drug users. Links between nodes represent drug affiliations. All links are reciprocated.

In this simulation study stratification is based on drug-use status and individual degree; those of individual degree less than three and which are non-injection drug users comprise the first stratum, those of individual degree less than three and which are injection drug users comprise the second stratum, and those of individual degree greater than or equal to three comprise the third stratum.

The Bernoulli sampling parameter is set to 0.10 for selection of the initial sample. A total of 5000 samples are selected; the average initial sample size is 59.68, first wave is 95.39, second wave is 107.25, and final sample is 262.33. The burn-in is set to 10\% for the Bayes data augmentation procedure. Gelman-Rubin statistics \citep{Gelman1992} are used to determine that sufficient lengths for the MCMC chains is 2000.
The prior for the population size is chosen to be flat, which corresponds to $a=0$ in Expression \ref{postNblock}, since values greater than zero will pull the Bayes estimate down and likely give rise to biased estimates.

Four other population size estimators are explored. First, solely for comparison purposes the two-sample bias-adjusted Lincoln-Petersen estimator \citep{Chapman1951, Seber1970} is used and is denoted as $\hat{N}_{LP}$. The samples are selected independently and completely at random where the size of one sample is equal to the expected size of the initial sample, and the size of the other sample is equal to the expected size of the first wave when applying the aforementioned one-wave snowball sampling design. Of course, selecting two samples at random and independently may be challenging when studying a hidden population. Hence, the following three Frank and Snijders' one-sample estimators \citep{Frank1994} are used. It is noted here that in the event that more than one wave is obtained for the sample, these three estimators are limited to the observations contained in the initial sample and to the first wave. In contrast, the Bayes strategy detailed in this paper has the added benefit of basing inference on the full sample observations.

1) The $\hat{N}_1$ moment-based estimator based on a Bernoulli ($\delta$) graph model and Bernoulli sample: Define $r$ to be the number of links in the initial sample and $s$ to be the number of links from the initial sample to the first wave. Conditional on $n_0$, moment estimators can be derived for $N$ and $\delta$, which lead to
\begin{align}
\hat{N}_1=\frac{n_0r+(n_0-1)s}{r}.
\end{align}
When conditioning on $n_0$, the statistics $r$ and $s$ are independent and each follow a binomial distribution. Based on a first order Taylor series, the asymptotic variance of $\hat{N}_1$ can be approximated by
\begin{align}
\hat{\text{var}}(\hat{N}_1)=\frac{(n_0^2-n_0-r)(n_0-1)s(s+r)}{n_0r^3}.
\end{align}

2) The $\hat{N}_3$ maximum likelihood estimator based on a Bernoulli graph model and Bernoulli sample: Define $t=r+s$. It can be shown that $n_1$ and $t$ are sufficient statistics corresponding to a graph likelihood based on counts of the number of nominations from the initial sample to each individual in the first wave. The maximum likelihood estimator can be shown to be that which satisfies
\begin{align}
1-\frac{n_1}{N-n_0} = \bigg[1-\frac{t}{n_0(N-1)}\bigg]^{n_0}.
\end{align}
An asymptotic approximation to the variance of this estimator is based on the assumption that, as $n_0$ and $N$ tend to infinity, the expected degree tends to a positive finite limit while $\frac{n_0}{N}$ and $\frac{n_0^2}{N}$ respectively tend to zero and infinity. It can be shown that the corresponding estimator is
\begin{align}
\hat{\text{var}}(\hat{N}_3) = \frac{(\hat{N}_3-n_0)^2}{(t-n_1)}.
\end{align}

3) The $\hat{N}_5$ design, moment-based estimator based on a Bernoulli sample: Define $k$ to be the number of nodes in the initial sample connected to at least one other node in the initial sample. The moment equations for $n_0$, $r$, and $k$ give the rise to the consistent estimator, based on the assumption that as $N$ and the Bernoulli sample parameter respectively tend to infinity and zero the initial sample size tends to infinity while all degrees are bounded, as
\begin{align}
\hat{N}_5=\frac{n_0k + (n_0-1)n_1}{k}.
\end{align}
A modified jackknife estimator is
\begin{align}
\hat{\text{var}}(\hat{N}_5)=\frac{(n_0-2)}{2n_0}\sum\limits_{i\epsilon S_0}(\hat{N}_{5(i)} - \hat{N}_{5(.)})^2.
\end{align}

Nominal 95\% confidence interval coverage rates corresponding to the other population size estimators are based on the central limit theorem (CLT). Coverage rates of the population size and MLEs of graph parameters based on a full graph realization are based on the equal-tailed 95\% credible interval from the posterior distribution of the Bayes estimates. Table \ref{Empirical_Population Size Estimators} presents the approximate expectation, variance (Var.), mean squared error (MSE),  coverage rate of corresponding confidence/probability intervals, and average length (Avg. Length) of the intervals for the population size estimators. Results corresponding to estimates of the model parameters are presented in the appendix.

\begin{longtable}{l*{10}{r}r}
\caption{Population size estimators for empirical study when initial sample is selected with a Bernoulli sampling design.}
\endfirsthead
\multicolumn{10}{l}
{{Table \ref{Empirical_Population Size Estimators} continued from previous page}} \\
  \hline
Estimator                               &Expectation     &Var.       &MSE        &Coverage Rate    &Avg. Length           \\\hline
\endhead
Estimator                               &Expectation     &Var.       &MSE        &Coverage Rate    &Avg. Length           \\\hline
$\hat{N}_{LP}$                          &599             &35,471     &35,487     &0.861            &644\\
$\hat{N}_1$                             &791             &366,667    &405,029    &0.866            &2,207\\
$\hat{N}_3$                             &349             &8,553      &69,215     &0.104            &186\\
$\hat{N}_5$                             &729             &233,075    &251,101    &0.936            &1,667 \\
$\hat{N}$ (Bayes, one wave)             &477             &10,452     &24,309     &0.614            &324 \\
$\hat{N}$ (Bayes, two waves)            &476             &2,291      &16,525     &0.368            &181 \\\hline
\label{Empirical_Population Size Estimators}
\end{longtable}
\vspace*{-\baselineskip}

Oftentimes, when sampling from a hidden population, it is not uncommon for the more central/conspicuous individuals in a population to be selected for the initial sample. We therefore explore the use of the new strategy when the initial sample is selected with probability proportional to individual degree plus one and the probabilities are scaled so the expected sample size is equal to the sampling parameter times the population size. Table \ref{Empirical_Population Size Estimators_unequal_selection} presents the approximate expectation for the population size estimators. Results corresponding to estimates of the model parameters are presented in the appendix. The average initial sample size is 59.60, first wave is 139.40, second wave is 72.83, and final sample is 271.83.
\begin{longtable}{l*{10}{r}r}
\caption{Population size estimators when selection for initial sample is with probability proportional to individual degree plus one, where probabilities are scaled so the expected sample size is equal to the sampling parameter times the population size.}
\endfirsthead
\multicolumn{10}{l}
{{Table \ref{Empirical_Population Size Estimators_unequal_selection} continued from previous page}} \\
  \hline
Estimator                               &Expectation     &Var.       &MSE        &Coverage Rate    &Avg. Length           \\\hline
\endhead
Estimator                               &Expectation     &Var.       &MSE        &Coverage Rate    &Avg. Length           \\\hline
$\hat{N}_{LP}$                          &598             &19,547     &19,558     &0.915            &508\\
$\hat{N}_1$                             &184             &745        &169,361    &0.000            &51\\
$\hat{N}_3$                             &223             &257        &138,819    &0.000            &38\\
$\hat{N}_5$                             &280             &945        &100,136    &0.000            &117 \\
$\hat{N}$ (Bayes, one wave)             &286             &714        &96,115     &0.000            &80 \\
$\hat{N}$ (Bayes, two waves)            &348             &333        &61,123     &0.000            &69  \\\hline
\label{Empirical_Population Size Estimators_unequal_selection}
\end{longtable}
\vspace*{-\baselineskip}

\section{Discussion}

In this paper we present a new strategy based on the multi-wave snowball sampling design, generalized stochastic block model, and Bayesian data augmentation procedure. We explore the procedure via simulation studies based on an empirical and several simulated populations, and demonstrate that our inference strategy has the potential to result in statistically efficient estimators for population attributes.

As demonstrated in the empirical and simulation studies, in terms of bias and variance, the Bayes data augmentation estimators for the population size often perform as well or better than the other estimators. With respect to the empirical study, the bias and MSE of the Bayes population size estimators are lowest amongst the one-sample estimators. With respect to the simulation studies, in the exception of a few cases, when sample sizes are reasonable the biases of the one-sample estimators are roughly on the same order of magnitude but the MSE is typically lowest for the Bayes population size estimators. When sample sizes are small, the bias of the two-wave Bayes population size estimate is amongst the lowest and the variance is substantially lower relative to the other one-sample estimators. Though the two-sample Lincoln-Petersen estimator typically has the least amount of bias amongst all estimators, the Bayes population size estimators have a substantially smaller variance and, in most cases, MSE. Hence, when heterogeneity exists in the network graph that can be accounted for with the stochastic block graph model, the new strategy has the ability to account for the clustering effects and therefore mitigate the bias and provide efficient estimates for the population size. Further, it provides the advantage of efficiently estimating the model parameters, which correspond to the proportion of individuals within each strata and propensity of links to arise within and between each and all pairs of strata.

The estimates corresponding to the empirical study are reasonable, and this suggests that there is a level of robustness of the strategy when applied to such networked populations. Some bias is evident in the stochastic block graph model parameter estimates. This is likely due to the erratic clustering effects, transitivity, and higher-order dependence that cannot be fully accounted for in the model with either the limited covariate information or strong reliance on the independence of links conditional on the identifiable strata memberships. Consequently, this affects the bias and coverage rates of the Bayes population size estimators. In comparison, the large standard errors of some of the other one-sample estimators offset their bias to give reasonable coverage rates. However, as remarked upon above, the Bayes population size estimators are noticeably less biased than the other one-sample estimators and this indicates that the strategy has much utility. With respect to variance reduction, the benefit of adding a second wave to the sample is noticeable.

For cases where the stochastic block model fits the network, the coverage rates perform well. Additionally, coverage rates based on two waves of sampling are superior when the initial sample size is small; there is a significant reduction in bias and variance of the Bayes population size estimators, variance of the $\underline{\lambda}$ estimators, and bias and variance of the $\underline{\beta}$ estimators. In cases where the model is based on a large number of parameters, some bias in the estimators is evident and this affects the coverage rates of the Bayes population size and model parameter estimators. In such cases, either larger sample sizes should be selected to obtain approximately unbiased estimators or the number of parameters should be reduced.

When the initial sample selection procedure selects the more central individuals in the population, a large degree of bias is introduced in all the estimators, which in turn affects the coverage rates. However, when the stochastic block model fits the population well and there are relatively few isolated individuals in the population, the benefits of using the Bayes estimators based on a two-wave snowball sample are noticeable in terms of all measures of performance. In general, when the conspicuous individuals are more likely to be selected for the initial sample the new approach has the ability to estimate characteristics related to the more central individuals in the population.

As the prior distribution on the population size moves away from a flat prior, the bias-variance tradeoff of the Bayes estimator for the population size is evident; the expected value of the resulting posterior distribution of the population size is pushed down while the variance is reduced. Consequently, the coverage rates are highest when the prior distribution is flat.

As mentioned in Section 2, the choice of stratification variables should be based on those which account for the pattern of social links between the members of the population. An additional factor to consider basing stratification on is the individual degree, as is done in the empirical study, as this can potentially account for clustering effects within the network. The inclusion of this variable in the empirical study significantly reduced the bias in the Bayes population size estimators. Hence, when there is an absence of covariate information, and the model does not adequately account for the clustering effects, it may be worthwhile to consider estimation based on such information as this may serve as a proxy for latent covariate information.

When sampling from a hidden population the assumption of drawing a Bernoulli sample is unlikely to hold. An approximate Bernoulli design may be assumed if selection is made across several sources of contact, and clusters of individuals within each source are avoided for selection. Even with such a strategy, an over-representation of the more central individuals can be expected, thereby biasing the population size estimators downwards. Also, there may be poorly-connected subsets of the network which are less accessible than others, which in-turn will present similar challenges for estimation. However, if a priori information is available for such subsets than this may be worked into the prior distribution for the $\underline{\lambda}$ and $\underline{\beta}$ parameters (see the discussion in subsection \ref{Posterior Distributions of Model Parameters}). Developments using a similar approach to that presented in this paper and with more elaborate designs, such as those based on initial sample selection probabilities which are a function of stratum membership and/or individual degree, would be useful for such complex and practical situations.

Inference based on a snowball sampling design is facilitated when covariate information is fully observed for the sampled individuals. This may be challenging when members of the target population possess a socially stigmatized or embarrassing behaviour. However, this challenge can be overcome as the methods presented in this study can be extended for the case when covariate information of some of the sampled individuals is not observed. Extensions to allow for relaxing the requirement of tracing all social links from members selected at intermediate waves should also be investigated. Indeed, such work could be highly beneficial since in an empirical setting some covariate information may not be observable and/or social links may not be traceable, such as when there is a chance that a potential recruit will become hostile during the recruitment process.


\bibliographystyle{biom}
\bibliography{MasterReferences}

\appendix
\section{Proof}

After sampling a population size $N$ from the corresponding posterior distribution presented in Expression (4), we now make use of the labels $1,2,...,N$ in the augmentation procedure solely for imputation purposes. We continue with the observed graph data labeled $d_{Obs'} = \{\text{S}, \underline{C}\unb{S},Y_{S\setminus S_W,U}\}$ where $U$ is a hypothetical population of size equal to that most recently sampled.

The probability distribution of $\underline{C}\ub{s}$ given $d_{Obs'}$ is obtained as
\begin{align}
  \tP(\underline{C}\ub{S} | d_{Obs'}) &= \tP(\underline{C}\ub{S} | \text{S}  , \underline{C}\unb{S}, Y_{ S\setminus S_W,U})  \notag \\
     &= \frac{\tP(\text{S}, \underline{C}\unb{S},  Y_{S\setminus S_W,U} | \underline{C}\ub{S})   \cdot   \tP(\underline{C}\ub{S})} {\tP(d_{Obs'})} \notag \\
     &= \frac{\tP(\text{S} | \underline{C}\unb{S}, Y_{S\setminus S_W,U}, \underline{C}\ub{S})   \cdot   \tP(\underline{C}\unb{S},  Y_{S\setminus S_W,U} | \underline{C}\ub{S})   \cdot   \tP(\underline{C}\ub{S})} {\tP(d_{Obs'})} \notag \\
     &= \frac{\tP(\text{S} | \underline{C}\unb{S}, Y_{S\setminus S_W,U})} {\tP(d_{Obs'})}   \cdot   \tP(Y_{S\setminus S_W,U} | \underline{C}\unb{S},  \underline{C}\ub{S})   \cdot   \tP(\underline{C}\unb{S},| \underline{C}\ub{S})   \cdot   \tP(\underline{C}\ub{S})\notag \\
     &= \frac{\tP(\text{S} | \underline{C}\unb{S},  Y_{S\setminus S_W,U})} {\tP(d_{Obs'})}   \cdot   \tP(Y_{S\setminus S_W,U} | \underline{C})   \cdot    \tP(\underline{C}).
\end{align}

\noindent We note here that $\underline{C}\ub{S}$ is dropped from the first term since the snowball sampling design, which is adaptive in nature, only depends on the information collected in the sample (see \citet{Thompson1996} for further details). We now have
\begin{align}
  \tP(\underline{C}\ub{S} | d_{Obs'}) &= \frac{\tP(\text{S} | \underline{C}\unb{S},  Y_{S\setminus S_W,U})} {\tP(d_{Obs'})}   \cdot   \tP(Y_{S\setminus S_W,U} | \underline{C})   \cdot       \prod_{i = 1}^N{\tP(C_i)}
\end{align}


\noindent where
\begin{align}
&P (Y_{S\setminus S_W,U} | \underline{C})= \prod\limits_{i = 1}^{n-n_W} \prod\limits_{\underset{i<j}{j=1:}}^{n} \tP (Y_{ij} | C_i, C_j) \cdot\prod\limits_{i = 1}^{n-n_W}\prod\limits_{k=n+1}^N \tP (Y_{ik} | C_i, C_k)\text{ .}
\end{align}

\noindent Therefore, by the factorization theorem we have
\begin{align}
  P(\underline{C}\ub{S} | d_{Obs'}) &= \prod_{i \in \bar{\text{S}}}   \tP(C_i | d_{Obs'}) \notag \\
  &= \prod_{i \in \bar{\text{S}}}   \tP(C_i | \text{S}, \underline{C}\unb{S}, Y_{S\setminus S_W,U})\notag \\
   &=  \prod_{i \in \bar{\text{S}}}   \tP(C_i | \text{S}, \underline{C}_{S\setminus S_W}, Y_{S\setminus S_W, i}) \text{ .}
\end{align}

Next, we take any $i \in \bar{\text{S}}$. Then, for any stratum $k = 1, 2, ... \, G$,
\begin{align}
&P(C_i = k | \text{S}, \underline{C}_{S\setminus S_W}, Y_{S\setminus S_W,i}) \notag \\
&= \frac{\tP(C_i = k, \text{S}, \underline{C}_{S\setminus S_W}, Y_{S\setminus S_W,i})}
   {\sum\limits_{\ell=1}^G \tP(C_i=\ell, \text{S}, \underline{C}_{S\setminus S_W},  Y_{S\setminus S_W,i})}\notag\\
&= \frac{\tP(C_i = k)   \cdot   \tP(Y_{S\setminus S_W , i} | \text{S}, \underline{C}_{S\setminus S_W}, C_i=k)}
       {\sum\limits_{\ell=1}^{G} \left[ \tP(C_i = \ell)   \cdot   \tP(Y_{S\setminus S_W,i} | \text{S}, \underline{C}_{S\setminus S_W},  C_i=\ell )  \right] }\notag \\
&= \frac{\tP(C_i = k)   \cdot   \prod\limits_{j=1}^{n-n_W}   \tP(Y_{ij}=0 | \text{S}, C_j,C_i=k)}
      {\sum\limits_{\ell=1}^{G}   \left[   \tP(C_i = \ell)   \cdot   \prod\limits_{j = 1}^{n-n_W}   \tP(Y_{ij}=0 | \text{S}, C_j,C_i=\ell )\right] } \notag \\
&=\frac{\lambda_k\prod\limits_{j=1}^{n-n_W}(1-\beta_{C_j,k})}
{\sum\limits_{\ell=1}^G\bigg(\lambda_l\prod\limits_{j=1}^{n-n_W}(1-\beta_{C_j,\ell})\bigg)} \text{ .}
\label{AA2appendix}
\end{align}

\clearpage

\section{Overview of Empirical and Simulation Studies}

The objective of the empirical and simulation studies are to 1) compare the performance of the Bayes estimator of the population size with the bias-adjusted Lincoln-Petersen estimator \citep{Chapman1951, Seber1970} and a set of Frank and Snijders' estimators \citep{Frank1994}; 2) conduct a sensitivity analysis of the prior distribution of the population size on the estimates of the population size and model parameters; the prior distributions considered are $\propto \frac{1}{N^a}$ where $a=0,1,2$, and the corresponding estimates are respectively denoted as $\hat{N}_{Bayes,0}$, $\hat{N}_{Bayes,1}$, and $\hat{N}_{Bayes,2}$; 3) gauge the improvement in estimators when additional waves are obtained for the sample; and 4) evaluate any bias in the estimators when selection for the initial sample is proportional to individual degree plus one.

All simulation studies are conducted in the R programming language \citep{Rprogram}. The network graphs are generated with the aid of the `igraph' package \citep{igraph}.

A summary and discussion of the simulation results can be found in the main paper.

\section{Empirical Study}

This section presents the Bayes data augmentation estimators for the model parameters from the empirical study, as presented in the main paper. Maximum likelihood estimates (MLEs) of the model parameters based on a full graph realization are $\hat{\underline{\lambda}}_{MLE}=(3.345E-01, 3.647E-01, 3.008E-01)$ and $\hat{\underline{\beta}}_{MLE}=(3.553E-04, 3.705E-04, 1.348E-03, 1.280E-03, 2.188E-03, 3.408E-02)$.

\begin{table}[H]
\caption{Bayes data augmentation estimators of the model parameters for empirical study based on one wave of sampling with selection of initial sample via a Bernoulli design.}
\centering
\begin{tabular}{rrrrrr}
  \hline
Estimator            &Mean   &Var.               &MSE                &Coverage Rate  &Avg. Length     \\\hline
  \hline
$\hat{\lambda}_1$   &   3.183E-01 & 2.618E-03 & 2.877E-03 & 9.410E-01 & 2.025E-01 \\
$\hat{\lambda}_2$   &   3.559E-01 & 2.460E-03 & 2.538E-03 & 9.486E-01 & 1.984E-01 \\
$\hat{\lambda}_3$   &   3.258E-01 & 1.951E-03 & 2.572E-03 & 9.338E-01 & 1.839E-01 \\
$\hat{\beta}_{1,1}$ &   1.013E-03 & 4.451E-07 & 8.770E-07 & 8.902E-01 & 2.790E-03 \\
$\hat{\beta}_{1,2}$ &   7.313E-04 & 1.365E-07 & 2.667E-07 & 8.914E-01 & 1.552E-03 \\
$\hat{\beta}_{1,3}$ &   2.029E-03 & 5.309E-07 & 9.954E-07 & 8.546E-01 & 2.799E-03 \\
$\hat{\beta}_{2,2}$ &   2.251E-03 & 1.021E-06 & 1.963E-06 & 8.904E-01 & 4.148E-03 \\
$\hat{\beta}_{2,3}$ &   3.104E-03 & 7.499E-07 & 1.588E-06 & 8.386E-01 & 3.470E-03 \\
$\hat{\beta}_{3,3}$ &   4.324E-02 & 1.266E-04 & 2.105E-04 & 6.236E-01 & 2.438E-02 \\
   \hline
   \label{Empirical_One_Wave}
\end{tabular}
\end{table}

\begin{table}[H]
\caption{Bayes data augmentation estimators of the model parameters for empirical study based on two waves of sampling with selection of initial sample via a Bernoulli design.}
\centering
\begin{tabular}{rrrrrr}
  \hline
Estimator            &Mean   &Var.               &MSE                &Coverage Rate  &Avg. Length     \\\hline
  \hline
$\hat{\lambda}_1$        & 2.973E-01 & 1.920E-03 & 3.299E-03 & 8.662E-01 & 1.751E-01 \\
$\hat{\lambda}_2$        & 3.514E-01 & 1.538E-03 & 1.715E-03 & 9.530E-01 & 1.675E-01 \\
$\hat{\lambda}_3$        & 3.513E-01 & 1.212E-03 & 3.757E-03 & 7.626E-01 & 1.524E-01 \\
$\hat{\beta}_{1,1}$      & 9.545E-04 & 2.382E-07 & 5.972E-07 & 8.514E-01 & 2.291E-03 \\
$\hat{\beta}_{1,2}$      & 6.463E-04 & 7.128E-08 & 1.473E-07 & 8.718E-01 & 1.157E-03 \\
$\hat{\beta}_{1,3}$      & 1.975E-03 & 2.098E-07 & 6.034E-07 & 7.758E-01 & 1.972E-03 \\
$\hat{\beta}_{2,2}$      & 1.747E-03 & 2.753E-07 & 4.931E-07 & 9.644E-01 & 2.643E-03 \\
$\hat{\beta}_{2,3}$      & 2.717E-03 & 2.249E-07 & 5.046E-07 & 8.888E-01 & 2.137E-03 \\
$\hat{\beta}_{3,3}$      & 4.104E-02 & 1.311E-05 & 6.149E-05 & 1.378E-01 & 7.936E-03 \\
   \hline
   \label{Empirical_Two_Waves}
\end{tabular}
\end{table}

\begin{table}[H]
\caption{Bayes data augmentation estimators of the model parameters for empirical study based on one wave of sampling with selection of initial sample proportional to node degree plus one.}
\centering
\begin{tabular}{rrrrrr}
  \hline
Estimator            &Mean   &Var.               &MSE                &Coverage Rate  &Avg. Length     \\\hline
  \hline
$\hat{\lambda}_1$     & 1.771E-01 & 1.335E-03 & 2.610E-02 & 3.940E-02 & 1.481E-01 \\
$\hat{\lambda}_2$     & 2.740E-01 & 1.482E-03 & 9.705E-03 & 4.604E-01 & 1.654E-01 \\
$\hat{\lambda}_3$     & 5.489E-01 & 1.676E-03 & 6.320E-02 & 0.000E+00 & 1.800E-01 \\
$\hat{\beta}_{1,1}$   & 6.916E-03 & 2.870E-05 & 7.174E-05 & 3.948E-01 & 2.018E-02 \\
$\hat{\beta}_{1,2}$   & 3.241E-03 & 2.934E-06 & 1.117E-05 & 3.374E-01 & 7.222E-03 \\
$\hat{\beta}_{1,3}$   & 5.473E-03 & 2.763E-06 & 1.978E-05 & 2.600E-02 & 6.416E-03 \\
$\hat{\beta}_{2,2}$   & 7.205E-03 & 1.033E-05 & 4.544E-05 & 2.444E-01 & 1.366E-02 \\
$\hat{\beta}_{2,3}$   & 6.459E-03 & 1.886E-06 & 2.013E-05 & 6.600E-03 & 5.756E-03 \\
$\hat{\beta}_{3,3}$   & 5.715E-02 & 3.771E-05 & 5.698E-04 & 6.000E-04 & 1.388E-02 \\
   \hline
   \label{Empirical_One_Wave_Unequal}
\end{tabular}
\end{table}

\begin{table}[H]
\caption{Bayes data augmentation estimators of the model parameters for empirical study based on two waves of sampling with selection of initial sample proportional to node degree plus one.}
\centering
\begin{tabular}{rrrrrr}
  \hline
Estimator            &Mean   &Var.               &MSE                &Coverage Rate  &Avg. Length     \\\hline
  \hline
$\hat{\lambda}_1$       & 1.855E-01 & 5.271E-04 & 2.270E-02 & 3.000E-03 & 1.226E-01 \\
$\hat{\lambda}_2$       & 3.193E-01 & 5.427E-04 & 2.604E-03 & 8.624E-01 & 1.396E-01 \\
$\hat{\lambda}_3$       & 4.952E-01 & 5.015E-04 & 3.826E-02 & 0.000E+00 & 1.411E-01 \\
$\hat{\beta}_{1,1}$     & 2.925E-03 & 1.425E-06 & 8.027E-06 & 3.632E-01 & 6.900E-03 \\
$\hat{\beta}_{1,2}$     & 1.224E-03 & 2.491E-07 & 9.778E-07 & 5.444E-01 & 2.279E-03 \\
$\hat{\beta}_{1,3}$     & 3.878E-03 & 3.304E-07 & 6.735E-06 & 4.000E-04 & 3.330E-03 \\
$\hat{\beta}_{2,2}$     & 2.743E-03 & 3.921E-07 & 2.532E-06 & 6.254E-01 & 3.945E-03 \\
$\hat{\beta}_{2,3}$     & 4.036E-03 & 1.743E-07 & 3.590E-06 & 1.060E-02 & 2.570E-03 \\
$\hat{\beta}_{3,3}$     & 3.617E-02 & 1.851E-06 & 6.198E-06 & 7.952E-01 & 6.192E-03 \\
   \hline
   \label{Empirical_Two_Wave_Unequal}
\end{tabular}
\end{table}

\section{Simulation Studies}

\subsection{Simulation Study 1}

The network population is generated from a two-strata stochastic block model. Parameters are set to $\lambda_1=\lambda_2=(1/2,1/2)$, $\beta_{11}=\beta_{12}=\beta_{22}=5/(N-1)$.

\subsubsection{Study 1.1: Bernoulli Initial Sample, $\mathbf{N=100}$}

\begin{figure}[H]
	\centering
\vspace{-1mm}
\centering
		
		\includegraphics[scale=0.4]{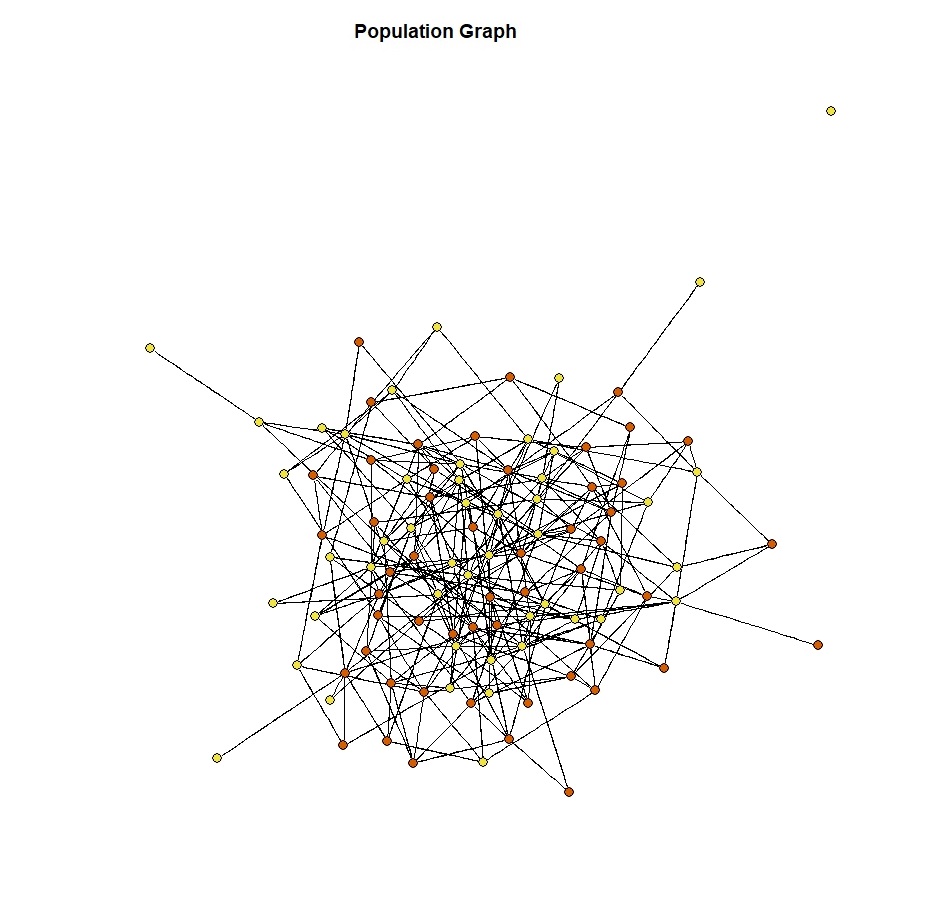}
\vspace{-3mm}
\caption{Population 1.1: Strata are represented by colour of nodes.}
\label{Pop1_1.jpeg}
\end{figure}

The sampling parameter is set to 0.20. The average initial sample size is 20.04, first wave is 53.33, and final sample is 73.38.

Gelman-Rubin statistics are based on chains of length 500, and 100 samples are selected to determine if the length of the chain is sufficient for estimating the population size and model parameters. The hyperparameter for the prior on the population size is set to zero. $\underline{\lambda}$ parameter seed values are set to (0.9,0.1) for the first chain and (0.1,0.9) for the second chain. $\underline{\beta}$ parameter seed values are set to 0.7 for the first chain and 0.3 for the second chain. The table below presents the mean and median of the corresponding Gelman-Rubin statistics.
\begin{table}[H]
\caption{Mean and median scores of Gelman-Rubin statistics corresponding with stochastic block model parameters for simulation study 1.1.}
\centering
\begin{tabular}{rrrrrrrr}
  \hline
Estimator            &Mean                  &Median                                   \\\hline
$\hat{N}$            &1.000                  &1.001                                \\
$\hat{\lambda}_1$    &1.001                  &1.000                                 \\
$\hat{\lambda}_2$    &1.001                  &1.000                                  \\
$\hat{\beta}_{1,1}$  &1.005                  &1.001   \\
$\hat{\beta}_{1,2}$  &1.004                  &1.002   \\
$\hat{\beta}_{2,2}$  &1.006                  &1.001   \\\hline
\label{1_1_GR}
\end{tabular}
\end{table}

For the Bayes data augmentation estimators the burn-in is set to 10\% and 5000 samples are selected. MLEs based on the full graph are: $\hat{\underline{\lambda}}_{MLE}=(4.800E-01, 5.200E-01)$ and $\hat{\underline{\beta}}_{MLE}=(6.028E-02, 5.449E-02 , 5.581E-02 )$. The tables below present the corresponding results from the simulation study.

\begin{table}[H]
\caption{Population size estimators for simulation study 1.1.}
\centering
\begin{tabular}{rrrrrrrr}
  \hline
Estimator            &Expectation       &Var.     &MSE        &Coverage Rate    &Avg. Length           \\\hline
$\hat{N}_{LP}$       &99.61             &375      &375        &0.921            &69.50\\
$\hat{N}_1$          &111.77            &2,697    &2,835      &0.877            &119.70\\
$\hat{N}_3$          &100.12            &128      &128        &0.974            &44.67\\
$\hat{N}_5$          &104.23            &1,162    &1,180      &0.934            &117.06 \\
$\hat{N}_{Bayes,0}$  &99.79             &91       &91         &0.935            &35.09\\
$\hat{N}_{Bayes,1}$  &99.23             &89       &90         &0.928            &34.49\\
$\hat{N}_{Bayes,2}$  &98.59             &84       &86         &0.930            &33.80\\\hline
\label{1_1_Pop_size_ests}
\end{tabular}
\end{table}

\begin{table}[H]
\caption{Bayes data augmentation estimators for simulation study 1.1. Population size is treated as known.}
\centering
\begin{tabular}{rrrrrrrr}
  \hline
Estimator            &Mean   &Var.    &MSE       &Coverage Rate  &Avg. Length     \\\hline
$\hat{\lambda}_1$    & 4.700E-01 & 1.392E-03 & 1.493E-03 & 9.962E-01 & 2.293E-01   \\
$\hat{\lambda}_2$    & 5.300E-01 & 1.392E-03 & 1.493E-03 & 9.962E-01 & 2.293E-01   \\
$\hat{\beta}_{1,1}$  & 6.545E-02 & 2.189E-04 & 2.456E-04 & 9.254E-01 & 5.337E-02   \\
$\hat{\beta}_{1,2}$  & 5.595E-02 & 4.864E-05 & 5.078E-05 & 9.698E-01 & 3.011E-02   \\
$\hat{\beta}_{2,2}$  & 5.699E-02 & 9.218E-05 & 9.359E-05 & 9.792E-01 & 4.457E-02   \\\hline
\label{1_1_N_known_results}
\end{tabular}
\end{table}

\begin{table}[H]
\caption{Bayes data augmentation estimators for simulation study 1.1. Prior 0 used for prior distribution on population size.}
\centering
\begin{tabular}{rrrrrrrr}
  \hline
Estimator            &Mean   &Var.    &MSE       &Coverage Rate  &Avg. Length     \\\hline
$\hat{\lambda}_1$     & 4.699E-01 & 1.292E-03 & 1.394E-03 & 9.984E-01 & 2.295E-01  \\
$\hat{\lambda}_2$     & 5.301E-01 & 1.292E-03 & 1.394E-03 & 9.984E-01 & 2.295E-01  \\
$\hat{\beta}_{1,1}$   & 6.668E-02 & 2.730E-04 & 3.139E-04 & 9.234E-01 & 5.855E-02  \\
$\hat{\beta}_{1,2}$   & 5.711E-02 & 8.568E-05 & 9.254E-05 & 9.528E-01 & 3.699E-02  \\
$\hat{\beta}_{2,2}$   & 5.833E-02 & 1.209E-04 & 1.273E-04 & 9.786E-01 & 5.015E-02  \\\hline
\label{1_1_prior_0_results}
\end{tabular}
\end{table}

\begin{table}[H]
\caption{Bayes data augmentation estimators for simulation study 1.1. Prior 1 used for prior distribution on population size.}
\centering
\begin{tabular}{rrrrrrrr}
  \hline
Estimator            &Mean   &Var.    &MSE       &Coverage Rate  &Avg. Length     \\\hline
$\hat{\lambda}_1$     & 4.699E-01 & 1.300E-03 & 1.402E-03 & 9.978E-01 & 2.294E-01   \\
$\hat{\lambda}_2$     & 5.301E-01 & 1.300E-03 & 1.402E-03 & 9.978E-01 & 2.294E-01   \\
$\hat{\beta}_{1,1}$   & 6.704E-02 & 3.382E-04 & 3.838E-04 & 9.250E-01 & 5.885E-02   \\
$\hat{\beta}_{1,2}$   & 5.739E-02 & 8.560E-05 & 9.400E-05 & 9.504E-01 & 3.707E-02   \\
$\hat{\beta}_{2,2}$   & 5.860E-02 & 1.251E-04 & 1.329E-04 & 9.766E-01 & 5.030E-02  \\\hline
\label{1_1_prior_1_results}
\end{tabular}
\end{table}

\begin{table}[H]
\caption{Bayes data augmentation estimators for simulation study 1.1. Prior 2 used for prior distribution on population size.}
\centering
\begin{tabular}{rrrrrrrr}
  \hline
Estimator            &Mean   &Var.    &MSE       &Coverage Rate  &Avg. Length     \\\hline
$\hat{\lambda}_1$     & 4.695E-01 & 1.310E-03 & 1.419E-03 & 9.972E-01 & 2.292E-01  \\
$\hat{\lambda}_2$     & 5.305E-01 & 1.310E-03 & 1.419E-03 & 9.972E-01 & 2.292E-01  \\
$\hat{\beta}_{1,1}$   & 6.772E-02 & 3.354E-04 & 3.908E-04 & 9.238E-01 & 5.924E-02  \\
$\hat{\beta}_{1,2}$   & 5.793E-02 & 8.522E-05 & 9.705E-05 & 9.508E-01 & 3.734E-02  \\
$\hat{\beta}_{2,2}$   & 5.870E-02 & 1.296E-04 & 1.379E-04 & 9.764E-01 & 5.039E-02  \\\hline
\label{1_1_prior_2_results}
\end{tabular}
\end{table}

\subsubsection{Study 1.2: Bernoulli Initial Sample, $\mathbf{N=1000}$}
\begin{figure}[H]
	\centering
\vspace{-1mm}
\centering
		
		\includegraphics[scale=0.4]{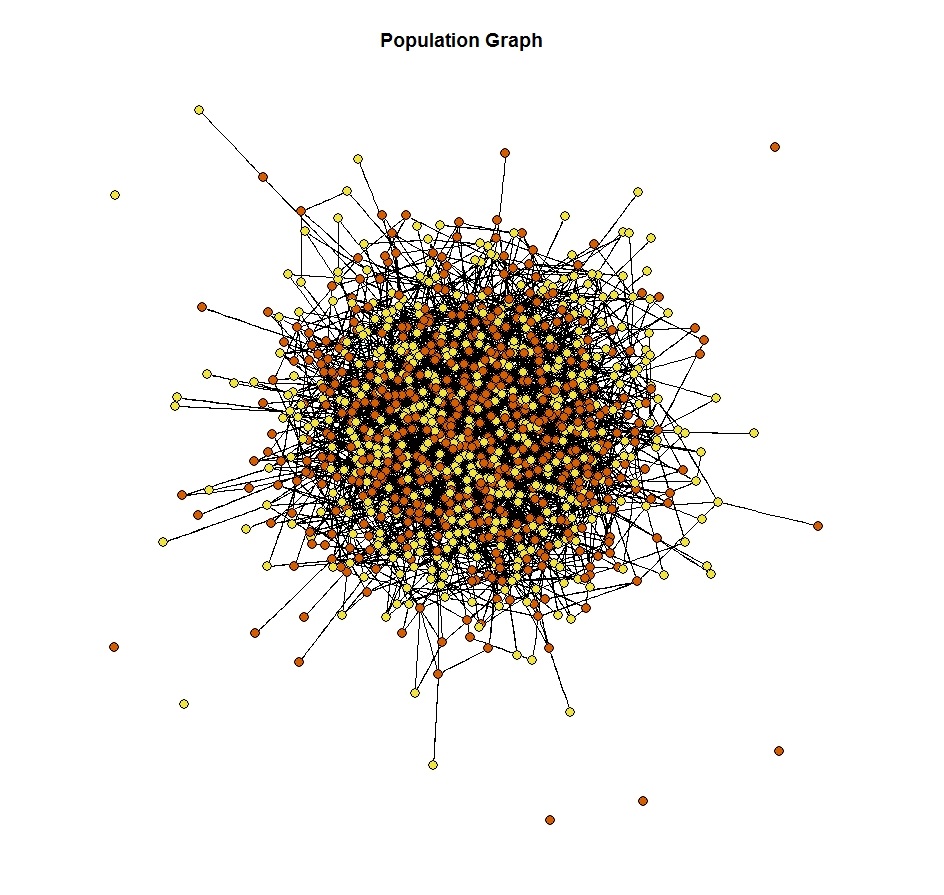}
\vspace{-3mm}
\caption{Population 1.2: Strata are represented by colour of nodes.}
\label{Pop1_2.jpeg}
\end{figure}

The sampling parameter is set to 0.06. The average initial sample size is 60.06, first wave is 240.75, and final sample is 300.81.

Gelman-Rubin statistics are based on chains of length 2000, and 100 samples are selected to determine if the length of the chain is sufficient for estimating the population size and model parameters. The hyperparameter for the prior on the population size is set to zero. $\underline{\lambda}$ parameter seed values are set to (0.9,0.1) for the first chain and (0.1,0.9) for the second chain. $\underline{\beta}$ parameter seed values are set to 0.7 for the first chain and 0.3 for the second chain. The table below presents the mean and median of the corresponding Gelman-Rubin statistics.
\begin{table}[H]
\caption{Mean and median scores of Gelman-Rubin statistics corresponding with stochastic block model parameters for simulation study 1.2.}
\centering
\begin{tabular}{rrrrrrrr}
  \hline
Estimator            &Mean                  &Median   \\\hline
$\hat{N}$            &1.018                 &1.006      \\
$\hat{\lambda}_1$    &1.005                 &1.002    \\
$\hat{\lambda}_2$    &1.005                 &1.002     \\
$\hat{\beta}_{1,1}$  &1.017                 &1.007   \\
$\hat{\beta}_{1,2}$  &1.016                 &1.007   \\
$\hat{\beta}_{2,2}$  &1.014                 &1.005   \\\hline
\label{1_2_GR}
\end{tabular}
\end{table}

For the Bayes data augmentation estimators the burn-in is set to 10\% and 5000 samples are selected. MLEs based on full graph are: $\hat{\underline{\lambda}}_{MLE}=(4.800E-02, 5.200E-02)$ and $\hat{\underline{\beta}}_{MLE}=(4.967E-03, 4.820E-03, 5.054E-03)$. The tables below present the corresponding results from the simulation study.

\begin{table}[H]
\caption{Population size estimators for simulation study 1.2.}
\centering
\begin{tabular}{rrrrrrrr}
  \hline
Estimator            &Mean      &Var.     &MSE        &Coverage Rate    &Avg. Length           \\\hline
$\hat{N}_{LP}$       &1,005     &56,852   &56,881     &0.902            &862\\
$\hat{N}_1$          &1,168     &412,514  &440,849    &0.875            &1,497 \\
$\hat{N}_3$          &1,028     &22,718   &23,517     &0.943            &538 \\
$\hat{N}_5$          &1,139     &328,286  &347,637    &0.940            &2,048 \\
$\hat{N}_{Bayes,0}$  &1,013     &19,197   &19,364     &0.933            &505 \\
$\hat{N}_{Bayes,1}$  &994       &17,894   &17,929     &0.926            &488 \\
$\hat{N}_{Bayes,2}$  &976       &16,542   &17,125     &0.915            &471 \\\hline
\label{1_2_Pop_size_ests}
\end{tabular}
\end{table}

\begin{table}[H]
\caption{Bayes data augmentation estimators for simulation study 1.2. Population size is treated as known.}
\centering
\begin{tabular}{rrrrrrrr}
  \hline
Estimator            &Mean   &Var.    &MSE       &Coverage Rate  &Avg. Length     \\\hline
$\hat{\lambda}_1$      & 4.786E-01 & 8.823E-04 & 8.842E-04 & 9.736E-01 & 1.323E-01   \\
$\hat{\lambda}_2$      & 5.214E-01 & 8.823E-04 & 8.842E-04 & 9.736E-01 & 1.323E-01   \\
$\hat{\beta}_{1,1}$    & 5.091E-03 & 4.799E-07 & 4.954E-07 & 9.556E-01 & 2.759E-03   \\
$\hat{\beta}_{1,2}$    & 4.870E-03 & 1.384E-07 & 1.409E-07 & 9.662E-01 & 1.605E-03   \\
$\hat{\beta}_{2,2}$    & 5.140E-03 & 3.961E-07 & 4.034E-07 & 9.544E-01 & 2.551E-03  \\\hline
\label{1_1_N_known_results}
\end{tabular}
\end{table}

\begin{table}[H]
\caption{Bayes data augmentation estimators for simulation study 1.2. Prior 0 used for prior distribution on population size.}
\centering
\begin{tabular}{rrrrrrrr}
  \hline
Estimator            &Mean   &Var.                   &MSE         &Coverage Rate  &Avg. Length     \\\hline

$\hat{\lambda}_1$      & 4.795E-01 & 8.744E-04 & 8.746E-04 & 9.744E-01 & 1.321E-01   \\
$\hat{\lambda}_2$      & 5.205E-01 & 8.744E-04 & 8.746E-04 & 9.744E-01 & 1.321E-01   \\
$\hat{\beta}_{1,1}$    & 5.208E-03 & 9.680E-07 & 1.026E-06 & 9.416E-01 & 3.801E-03   \\
$\hat{\beta}_{1,2}$    & 4.987E-03 & 5.653E-07 & 5.934E-07 & 9.412E-01 & 2.938E-03   \\
$\hat{\beta}_{2,2}$    & 5.272E-03 & 9.011E-07 & 9.485E-07 & 9.416E-01 & 3.664E-03   \\\hline
\label{1_2_prior_0_results}
\end{tabular}
\end{table}

\begin{table}[H]
\caption{Bayes data augmentation estimators for simulation study 1.2. Prior 1 used for prior distribution on population size.}
\centering
\begin{tabular}{rrrrrrrr}
  \hline
Estimator            &Mean   &Var.                   &MSE            &Coverage Rate  &Avg. Length     \\\hline

$\hat{\lambda}_1$      & 4.793E-01 & 8.651E-04 & 8.656E-04 & 9.760E-01 & 1.318E-01 \\
$\hat{\lambda}_2$      & 5.207E-01 & 8.651E-04 & 8.656E-04 & 9.760E-01 & 1.318E-01 \\
$\hat{\beta}_{1,1}$    & 5.311E-03 & 1.036E-06 & 1.154E-06 & 9.312E-01 & 3.844E-03 \\
$\hat{\beta}_{1,2}$    & 5.075E-03 & 5.640E-07 & 6.291E-07 & 9.328E-01 & 2.956E-03 \\
$\hat{\beta}_{2,2}$    & 5.344E-03 & 8.870E-07 & 9.709E-07 & 9.424E-01 & 3.688E-03 \\ \hline
\label{1_2_prior_1_results}
\end{tabular}
\end{table}

\begin{table}[H]
\caption{Bayes data augmentation estimators for simulation study 1.2. Prior 2 used for prior distribution on population size.}
\centering
\begin{tabular}{rrrrrrrr}
  \hline
Estimator            &Mean   &Var.                   &MSE            &Coverage Rate  &Avg. Length     \\\hline

$\hat{\lambda}_1$     & 4.794E-01 & 8.621E-04 & 8.625E-04 & 9.728E-01 & 1.312E-01 \\
$\hat{\lambda}_2$     & 5.206E-01 & 8.621E-04 & 8.625E-04 & 9.728E-01 & 1.312E-01 \\
$\hat{\beta}_{1,1}$   & 5.407E-03 & 1.014E-06 & 1.207E-06 & 9.312E-01 & 3.886E-03 \\
$\hat{\beta}_{1,2}$   & 5.166E-03 & 5.797E-07 & 6.995E-07 & 9.280E-01 & 2.985E-03 \\
$\hat{\beta}_{2,2}$   & 5.465E-03 & 9.251E-07 & 1.094E-06 & 9.298E-01 & 3.742E-03 \\ \hline
\label{1_2_prior_2_results}
\end{tabular}
\end{table}

\subsubsection{Study 1.3: Unequal Probability Initial Sample, $\mathbf{N=1000}$}
\begin{figure}[H]
	\centering
\vspace{-1mm}
\centering
		
		\includegraphics[scale=0.4]{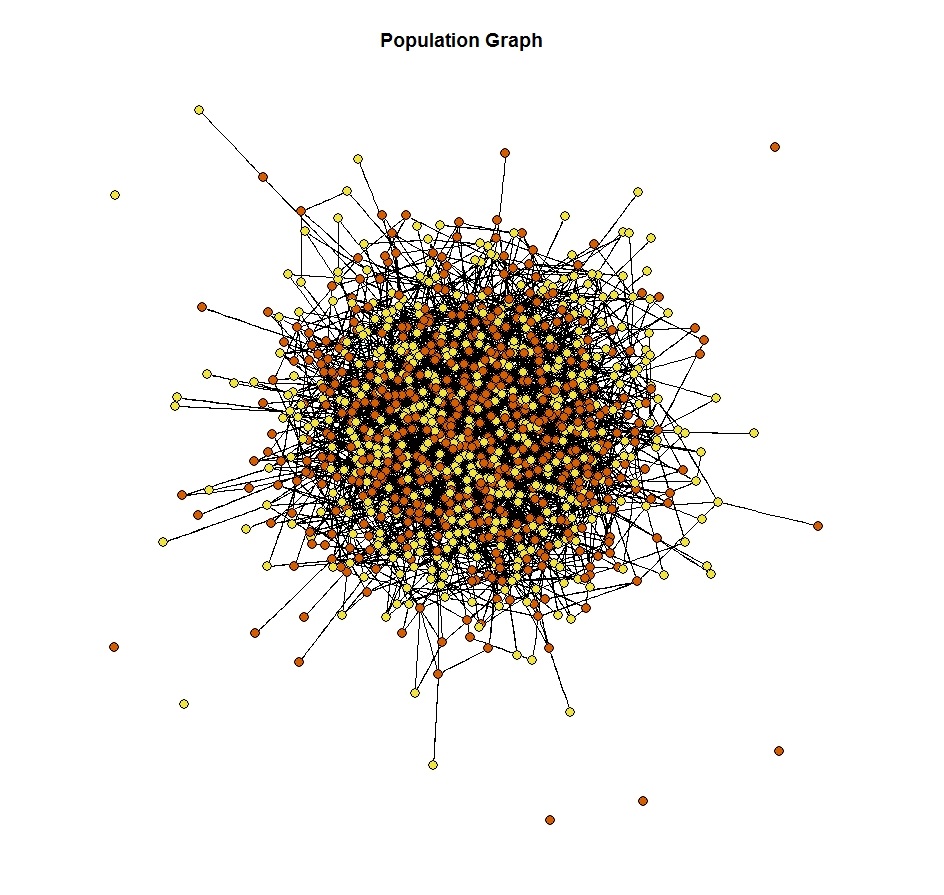}
\vspace{-3mm}
\caption{Population 1.3: Strata are represented by colour of nodes.}
\label{Pop1_3.jpeg}
\end{figure}

Selection for the initial sample is with probability proportional to individual degree plus one where probabilities are scaled so the expected size of the initial sample is equal to the sampling parameter times the population size. The sampling parameter is set to 0.06. The average initial sample size is 60.11, first wave is 271.15, and final sample is 331.25.

For the Bayes data augmentation estimators the burn-in is set to 10\% and 5000 samples are selected. MLEs based on full graph are: $\hat{\underline{\lambda}}_{MLE}=(4.800E-02, 5.200E-02)$ and $\hat{\underline{\beta}}_{MLE}=(4.967E-03, 4.820E-03,  5.054E-03)$. The tables below present the corresponding results from the simulation study. The prior for the population size is flat (prior 0).

\begin{table}[H]
\caption{Population size estimators for simulation study 1.3.}
\centering
\begin{tabular}{rrrrrrrr}
  \hline
Estimator            &Mean      &Var.     &MSE        &Coverage Rate    &Avg. Length           \\\hline
$\hat{N}_{LP}$       &1,002     &46,673   &46,679     &0.905            &792\\
$\hat{N}_1$          &953       &185,495  &187,672    &0.746            &947 \\
$\hat{N}_3$          &944       &14,951   &14,986     &0.930            &451 \\
$\hat{N}_5$          &952       &141,808  &144,150    &0.862            &1,228 \\
$\hat{N}_{Bayes,0}$  &998       &13,729   &13,734     &0.935            &433\\\hline
\label{1_3_Pop_size}
\end{tabular}
\end{table}

\begin{table}[H]
\caption{Bayes data augmentation estimators for simulation study 1.3.}
\centering
\begin{tabular}{rrrrrrrr}
  \hline
Estimator            &Mean   &Var.                   &MSE         &Coverage Rate  &Avg. Length     \\\hline
$\hat{\lambda}_1$      & 4.742E-01 & 7.108E-04 & 7.449E-04 & 9.796E-01 & 1.254E-01 \\
$\hat{\lambda}_2$      & 5.258E-01 & 7.108E-04 & 7.449E-04 & 9.796E-01 & 1.254E-01 \\
$\hat{\beta}_{1,1}$    & 6.184E-03 & 1.043E-06 & 2.524E-06 & 7.842E-01 & 4.085E-03 \\
$\hat{\beta}_{1,2}$    & 5.733E-03 & 5.620E-07 & 1.397E-06 & 7.732E-01 & 3.018E-03 \\
$\hat{\beta}_{2,2}$    & 6.061E-03 & 9.139E-07 & 1.928E-06 & 8.236E-01 & 3.807E-03 \\\hline
\label{1_3_prior_0_results}
\end{tabular}
\end{table}

\subsection{Simulation Study 2}

The network population is generated from a two-strata stochastic block model. Parameters are set to $\lambda_1=\lambda_2=(1/2,1/2)$, $\beta_{11}=12/(N-2), \beta_{12}=1/(N-1)$, and $\beta_{22}=6/(N-2)$.

\subsubsection{Study 2.1: Bernoulli Initial Sample, $\mathbf{N=100}$}

\begin{figure}[H]
	\centering
\vspace{-1mm}
\centering
		
		\includegraphics[scale=0.4]{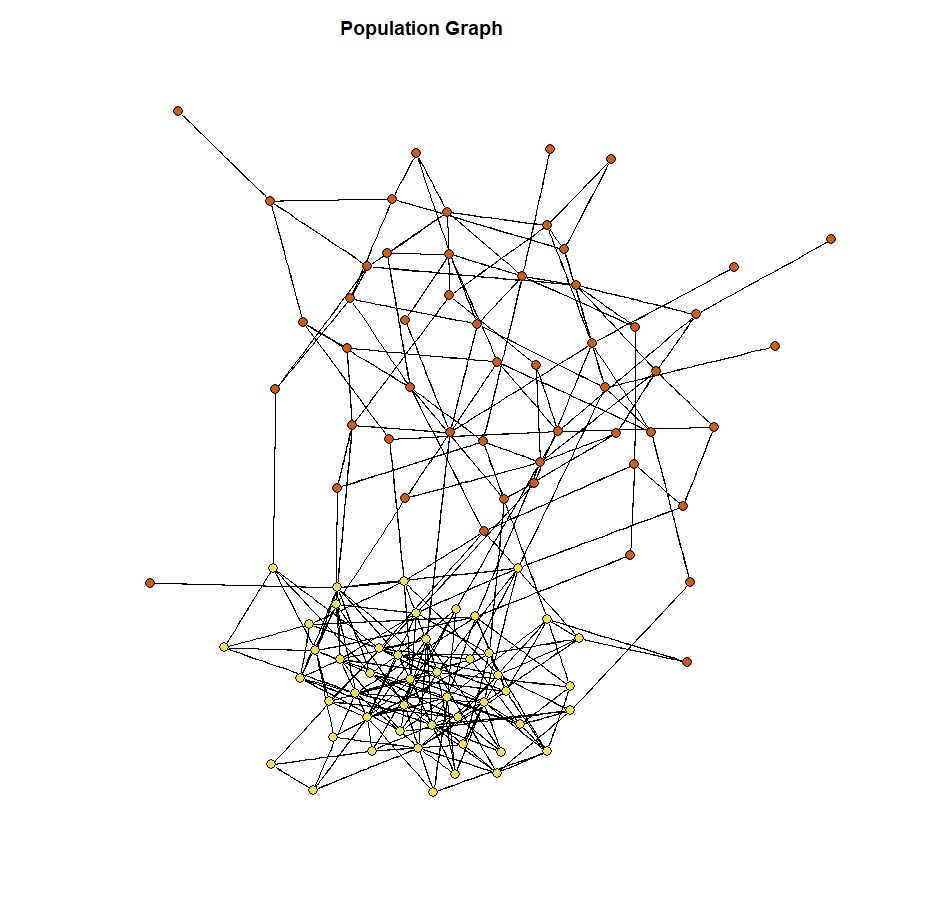}
\vspace{-3mm}
\caption{Population 2.1: Strata are represented by colour of nodes.}
\label{Pop2_1.jpeg}
\end{figure}

The sampling parameter is set to 0.20. The average initial sample size is 20.01, first wave is 51.77, and final sample is 71.78.

Gelman-Rubin statistics are based on chains of length 500, and 100 samples are selected to determine if the length of the chain is sufficient for estimating the population size and model parameters. The hyperparameter for the prior on the population size is set to zero. $\underline{\lambda}$ parameter seed values are set to (0.9,0.1) for the first chain and (0.1,0.9) for the second chain. $\underline{\beta}$ parameter seed values are set to 0.7 for the first chain and 0.3 for the second chain. The table below presents the mean and median of the corresponding Gelman-Rubin statistics.
\begin{table}[H]
\caption{Mean and median scores of Gelman-Rubin statistics corresponding with stochastic block model parameters for simulation study 2.1.}
\centering
\begin{tabular}{rrrrrrrr}
  \hline
Estimator            &Mean                  &Median                                   \\\hline
$\hat{N}$            &1.006                 &1.003                               \\
$\hat{\lambda}_1$    &1.005                 &1.005                                 \\
$\hat{\lambda}_2$    &1.001                 &1.001                                  \\
$\hat{\beta}_{1,1}$  &1.011                 &1.004   \\
$\hat{\beta}_{1,2}$  &1.004                 &1.001   \\
$\hat{\beta}_{2,2}$  &1.008                 &1.003   \\\hline
\label{2_1_GR}
\end{tabular}
\end{table}

For the Bayes data augmentation estimators the burn-in is set to 10\% and 5000 samples are selected. MLEs based on full graph are: $\hat{\underline{\lambda}}_{MLE}=(4.800E-01, 5.200E-01)$ and $\hat{\underline{\beta}}_{MLE}=(1.410E-01, 8.413E-03, 6.561E-02)$. The tables below present the corresponding results from the simulation study.

\begin{table}[H]
\caption{Population size estimators for simulation study 2.1.}
\centering
\begin{tabular}{rrrrrrrr}
  \hline
Estimator            &Mean      &Var.     &MSE        &Coverage Rate    &Avg. Length           \\\hline
$\hat{N}_{LP}$       &99.58     &390      &390        &0.848            &70.83\\
$\hat{N}_1$          &112.94    &3,017    &3,184      &0.882            &122.20\\
$\hat{N}_3$          &99.33     &143      &144        &0.928            &45.54\\
$\hat{N}_5$          &104.56    &1,289    &1,310      &0.940            &126.05 \\
$\hat{N}_{Bayes,0}$  &102.22    &131      &136        &0.940            &42.71 \\
$\hat{N}_{Bayes,1}$  &100.95    &120      &121        &0.942            &40.51 \\
$\hat{N}_{Bayes,2}$  &99.62     &109      &109        &0.934            &38.78 \\\hline
\label{2_1_Pop_size_ests}
\end{tabular}
\end{table}

\begin{table}[H]
\caption{Bayes data augmentation estimators for simulation study 2.1. Population size is treated as known.}
\centering
\begin{tabular}{rrrrrrrr}
  \hline
Estimator            &Mean   &Var.     &MSE        &Coverage Rate  &Avg. Length     \\\hline
$\hat{\lambda}_1$   &4.827E-01 & 1.563E-03 & 1.570E-03 & 9.940E-01 & 2.341E-01 \\
$\hat{\lambda}_2$   &5.173E-01 & 1.563E-03 & 1.570E-03 & 9.940E-01 & 2.341E-01 \\
$\hat{\beta}_{1,1}$ &1.441E-01 & 4.395E-04 & 4.496E-04 & 9.730E-01 & 8.279E-02 \\
$\hat{\beta}_{1,2}$ &9.613E-03 & 8.250E-06 & 9.688E-06 & 9.732E-01 & 1.250E-02 \\
$\hat{\beta}_{2,2}$ &6.866E-02 & 1.323E-04 & 1.417E-04 & 9.748E-01 & 5.099E-02 \\ \hline
\label{2_1_N_known_results}
\end{tabular}
\end{table}

\begin{table}[H]
\caption{Bayes data augmentation estimators for simulation study 2.1. Prior 0 used for prior distribution on population size.}
\centering
\begin{tabular}{rrrrrrrr}
  \hline
Estimator            &Mean   &Var.    &MSE       &Coverage Rate  &Avg. Length     \\\hline
$\hat{\lambda}_1$    &4.824E-01 & 2.329E-03 & 2.335E-03 & 9.896E-01 & 2.528E-01 \\
$\hat{\lambda}_2$    &5.176E-01 & 2.329E-03 & 2.335E-03 & 9.896E-01 & 2.528E-01 \\
$\hat{\beta}_{1,1}$  &1.439E-01 & 5.102E-04 & 5.188E-04 & 9.714E-01 & 8.832E-02 \\
$\hat{\beta}_{1,2}$  &9.710E-03 & 9.125E-06 & 1.081E-05 & 9.760E-01 & 1.337E-02 \\
$\hat{\beta}_{2,2}$  &6.991E-02 & 2.782E-04 & 2.967E-04 & 9.682E-01 & 6.673E-02 \\\hline
\label{2_1_prior_0_results}
\end{tabular}
\end{table}

\begin{table}[H]
\caption{Bayes data augmentation estimators for simulation study 2.1. Prior 1 used for prior distribution on population size.}
\centering
\begin{tabular}{rrrrrrrr}
  \hline
Estimator            &Mean   &Var.    &MSE       &Coverage Rate  &Avg. Length     \\\hline
$\hat{\lambda}_1$    &4.842E-01 & 2.199E-03 & 2.216E-03 & 9.906E-01 & 2.500E-01 \\
$\hat{\lambda}_2$    &5.158E-01 & 2.199E-03 & 2.216E-03 & 9.906E-01 & 2.500E-01 \\
$\hat{\beta}_{1,1}$  &1.448E-01 & 4.829E-04 & 4.980E-04 & 9.722E-01 & 8.807E-02 \\
$\hat{\beta}_{1,2}$  &9.850E-03 & 9.544E-06 & 1.161E-05 & 9.690E-01 & 1.351E-02 \\
$\hat{\beta}_{2,2}$  &7.111E-02 & 2.802E-04 & 3.105E-04 & 9.702E-01 & 6.678E-02 \\\hline
\label{2_1_prior_1_results}
\end{tabular}
\end{table}

\begin{table}[H]
\caption{Bayes data augmentation estimators for simulation study 2.1. Prior 2 used for prior distribution on population size.}
\centering
\begin{tabular}{rrrrrrrr}
  \hline
Estimator            &Mean   &Var.    &MSE       &Coverage Rate  &Avg. Length     \\\hline
$\hat{\lambda}_1$    &4.872E-01 & 2.184E-03 & 2.236E-03 & 9.898E-01 & 2.489E-01 \\
$\hat{\lambda}_2$    &5.128E-01 & 2.184E-03 & 2.236E-03 & 9.898E-01 & 2.489E-01 \\
$\hat{\beta}_{1,1}$  &1.460E-01 & 4.626E-04 & 4.884E-04 & 9.714E-01 & 8.826E-02 \\
$\hat{\beta}_{1,2}$  &9.940E-03 & 9.596E-06 & 1.193E-05 & 9.680E-01 & 1.366E-02 \\
$\hat{\beta}_{2,2}$  &7.260E-02 & 2.913E-04 & 3.402E-04 & 9.658E-01 & 6.767E-02 \\\hline
\label{2_1_prior_2_results}
\end{tabular}
\end{table}

\subsubsection{Study 2.2: Bernoulli Initial Sample, $\mathbf{N=1000}$}
\begin{figure}[H]
	\centering
\vspace{-1mm}
\centering
		
		\includegraphics[scale=0.4]{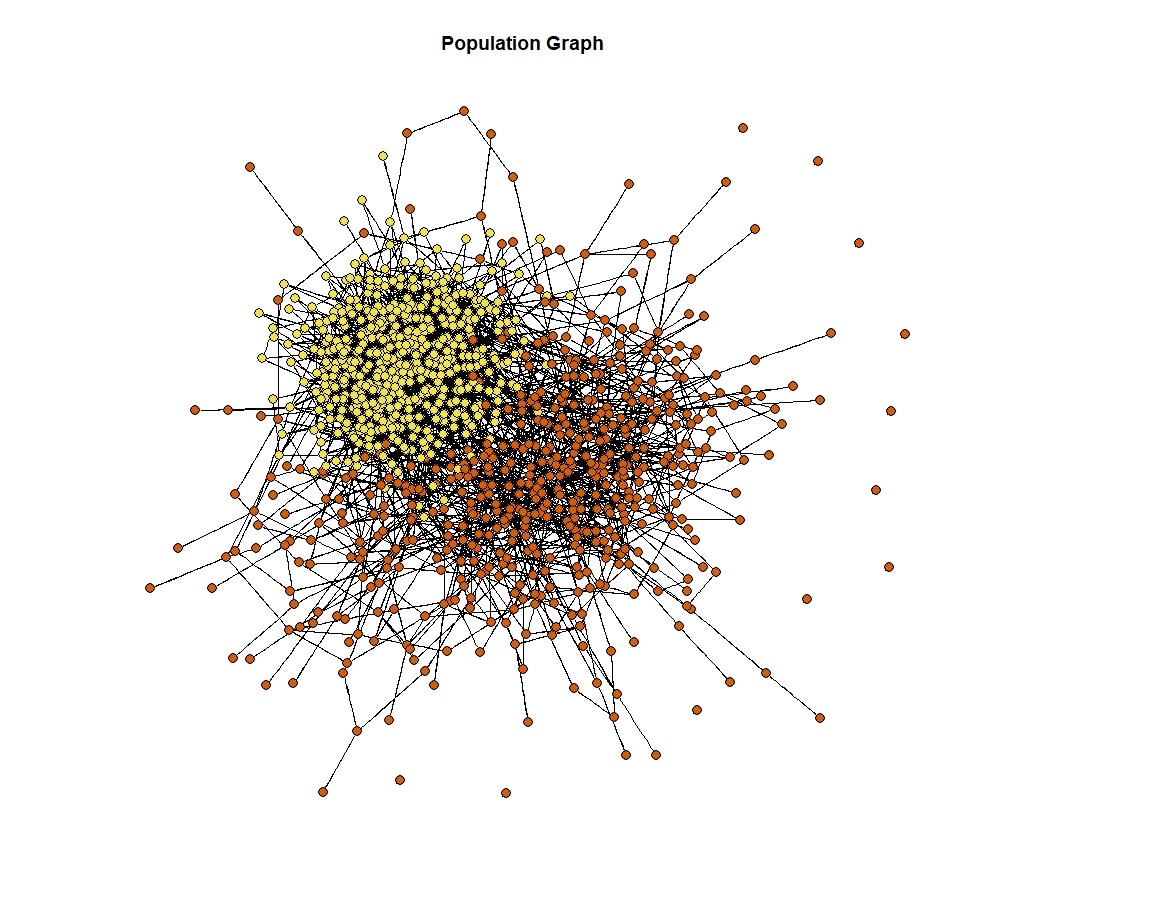}
\vspace{-3mm}
\caption{Population 2.2: Strata are represented by colour of nodes.}
\label{Pop2_2.jpeg}
\end{figure}

The sampling parameter is set to 0.06. The average initial sample size is 60.04, first wave is 235.83, and final sample is 295.86.

Gelman-Rubin statistics are based on chains of length 2000, and 100 samples are selected to determine if the length of the chain is sufficient for estimating the population size and model parameters. The hyperparameter for the prior on the population size is set to zero. $\underline{\lambda}$ parameter seed values are set to (0.9,0.1) for the first chain and (0.1,0.9) for the second chain. $\underline{\beta}$ parameter seed values are set to 0.7 for the first chain and 0.3 for the second chain. The table below presents the mean and median of the corresponding Gelman-Rubin statistics.
\begin{table}[H]
\caption{Mean and median scores of Gelman-Rubin statistics corresponding with stochastic block model parameters for simulation study 2.2.}
\centering
\begin{tabular}{rrrrrrrr}
  \hline
Estimator            &Mean                   &Median                                   \\\hline
$\hat{N}$            &1.022                  &1.009                                \\
$\hat{\lambda}_1$    &1.013                  &1.006                                 \\
$\hat{\lambda}_2$    &1.013                  &1.006                                  \\
$\hat{\beta}_{1,1}$  &1.021                  &1.010   \\
$\hat{\beta}_{1,2}$  &1.011                  &1.005   \\
$\hat{\beta}_{2,2}$  &1.024                  &1.008   \\\hline
\label{2_2_GR}
\end{tabular}
\end{table}

For the Bayes data augmentation estimators the burn-in is set to 10\% and 5000 samples are selected. MLEs based on full graph are: $\hat{\underline{\lambda}}_{MLE}=(4.800E-01, 5.200E-01)$ and $\hat{\underline{\beta}}_{MLE}=(1.193E-02, 9.696E-04, 6.025E-03)$. The tables below present the corresponding results from the simulation study.

\begin{table}[H]
\caption{Population size estimators for simulation study 2.2.}
\centering
\begin{tabular}{rrrrrrrr}
  \hline
Estimator            &Mean              &Var.     &MSE        &Coverage Rate    &Avg. Length           \\\hline
$\hat{N}_{LP}$       &1,005             &58,170   &58,199     &0.915            &872\\
$\hat{N}_1$          &1,173             &441,847  &471,697    &0.876            &1,496 \\
$\hat{N}_3$          &988               &22,137   &22,283     &0.895            &513 \\
$\hat{N}_5$          &1,137             &332,523  &351,341    &0.938            &2,081 \\
$\hat{N}_{Bayes,0}$  &1,009             &22,238   &22,317     &0.922            &519 \\
$\hat{N}_{Bayes,1}$  &991               &20,572   &20,657     &0.904            &501 \\
$\hat{N}_{Bayes,2}$  &971               &17,188   &18,003     &0.911            &483 \\\hline
\label{2_2_Pop_size_ests}
\end{tabular}
\end{table}

\begin{table}[H]
\caption{Bayes data augmentation estimators for simulation study 2.2. Population size is treated as known.}
\centering
\begin{tabular}{rrrrrrrrrrr}
  \hline
Estimator            &Mean   &Var.                   &MSE         &Coverage Rate  &Avg. Length     \\\hline
$\hat{\lambda}_1$    & 4.812E-01 & 1.714E-03 & 1.715E-03 & 9.510E-01 & 1.631E-01 \\
$\hat{\lambda}_2$    & 5.188E-01 & 1.714E-03 & 1.715E-03 & 9.510E-01 & 1.631E-01 \\
$\hat{\beta}_{1,1}$  & 1.215E-02 & 1.996E-06 & 2.044E-06 & 9.456E-01 & 5.495E-03 \\
$\hat{\beta}_{1,2}$  & 1.008E-03 & 3.264E-08 & 3.413E-08 & 9.480E-01 & 7.249E-04 \\
$\hat{\beta}_{2,2}$  & 6.187E-03 & 6.569E-07 & 6.831E-07 & 9.482E-01 & 3.150E-03 \\\hline
\label{2_2 N known_results}
\end{tabular}
\end{table}

\begin{table}[H]
\caption{Bayes data augmentation estimators for simulation study 2.2. Prior 0 used for prior distribution on population size.}
\centering
\begin{tabular}{rrrrrrrr}
  \hline
Estimator            &Mean   &Var.                   &MSE         &Coverage Rate  &Avg. Length     \\\hline
$\hat{\lambda}_1$   & 4.836E-01 & 1.875E-03 & 1.888E-03 & 9.458E-01 & 1.692E-01 \\
$\hat{\lambda}_2$   & 5.164E-01 & 1.875E-03 & 1.888E-03 & 9.458E-01 & 1.692E-01 \\
$\hat{\beta}_{1,1}$ & 1.239E-02 & 4.094E-06 & 4.308E-06 & 9.310E-01 & 7.427E-03 \\
$\hat{\beta}_{1,2}$ & 1.039E-03 & 6.033E-08 & 6.510E-08 & 9.330E-01 & 9.191E-04 \\
$\hat{\beta}_{2,2}$ & 6.446E-03 & 1.852E-06 & 2.030E-06 & 9.298E-01 & 5.017E-03 \\ \hline
\label{2_2_prior_0_results}
\end{tabular}
\end{table}

\begin{table}[H]
\caption{Bayes data augmentation estimators for simulation study 2.2. Prior 1 used for prior distribution on population size.}
\centering
\begin{tabular}{rrrrrrrr}
  \hline
Estimator            &Mean   &Var.                   &MSE            &Coverage Rate  &Avg. Length     \\\hline
$\hat{\lambda}_1$      & 4.845E-01 & 1.912E-03 & 1.932E-03 & 9.426E-01 & 1.677E-01 \\
$\hat{\lambda}_2$      & 5.155E-01 & 1.912E-03 & 1.932E-03 & 9.426E-01 & 1.677E-01 \\
$\hat{\beta}_{1,1}$    & 1.258E-02 & 4.072E-06 & 4.495E-06 & 9.268E-01 & 7.491E-03 \\
$\hat{\beta}_{1,2}$    & 1.068E-03 & 6.133E-08 & 7.094E-08 & 9.286E-01 & 9.351E-04 \\
$\hat{\beta}_{2,2}$    & 6.594E-03 & 1.957E-06 & 2.282E-06 & 9.122E-01 & 5.083E-03 \\ \hline
\label{2_2_prior_1_results}
\end{tabular}
\end{table}

\begin{table}[H]
\caption{Bayes data augmentation estimators for simulation study 2.2. Prior 2 used for prior distribution on population size.}
\centering
\begin{tabular}{rrrrrrrr}
  \hline
Estimator            &Mean   &Var.                   &MSE            &Coverage Rate  &Avg. Length     \\\hline
$\hat{\lambda}_1$    & 4.873E-01 & 1.796E-03 & 1.849E-03 & 9.460E-01 & 1.667E-01 \\
$\hat{\lambda}_2$    & 5.127E-01 & 1.796E-03 & 1.849E-03 & 9.460E-01 & 1.667E-01 \\
$\hat{\beta}_{1,1}$  & 1.274E-02 & 3.945E-06 & 4.612E-06 & 9.272E-01 & 7.492E-03 \\
$\hat{\beta}_{1,2}$  & 1.076E-03 & 5.974E-08 & 7.112E-08 & 9.258E-01 & 9.376E-04 \\
$\hat{\beta}_{2,2}$  & 6.721E-03 & 1.821E-06 & 2.304E-06 & 9.178E-01 & 5.140E-03 \\\hline
\label{2_2_prior_2_results}
\end{tabular}
\end{table}

\subsubsection{Study 2.3: Unequal Probability Initial Sample, $\mathbf{N=1000}$}
\begin{figure}[H]
	\centering
\vspace{-1mm}
\centering
		
		\includegraphics[scale=0.4]{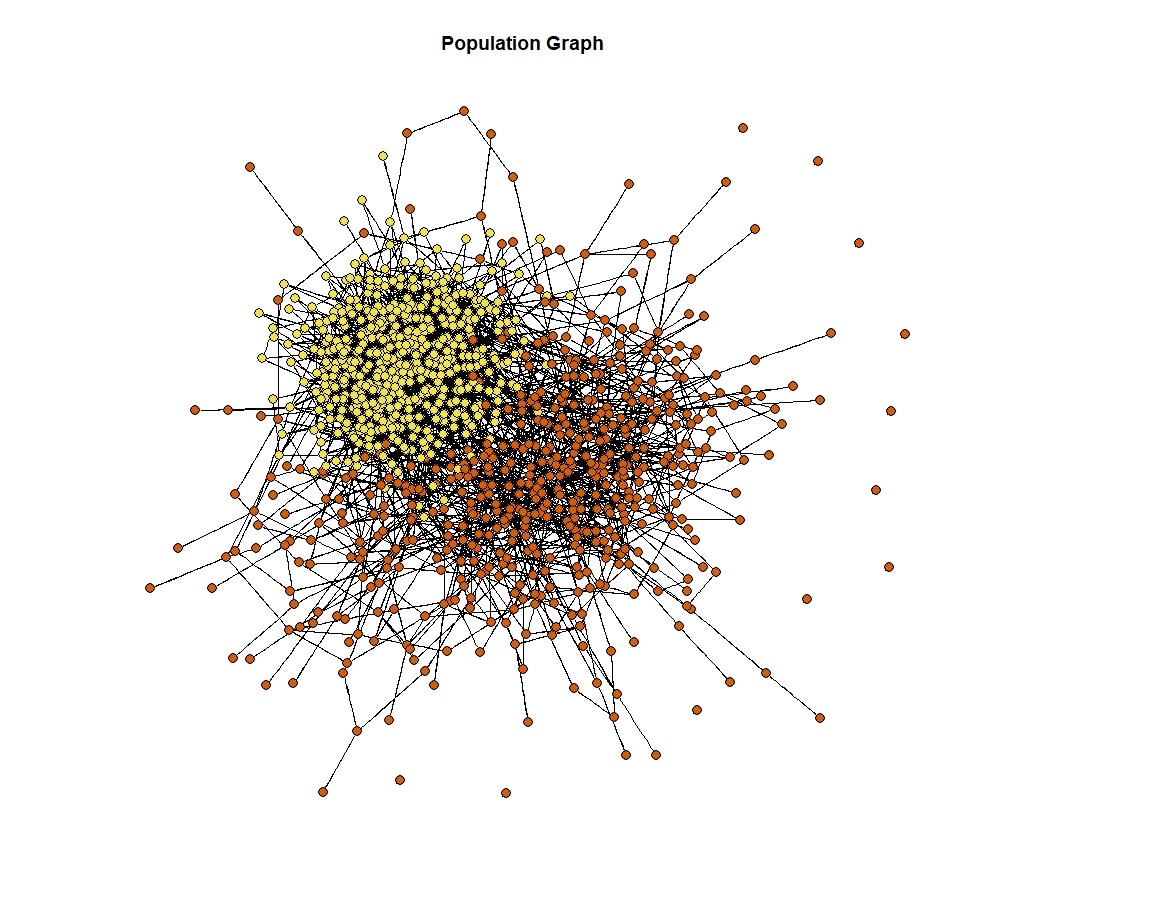}
\vspace{-3mm}
\caption{Population 2.3: Strata are represented by colour of nodes.}
\label{Pop2_2.jpeg}
\end{figure}

Selection for the initial sample is with probability proportional to individual degree plus one where probabilities are scaled so the expected size of the initial sample is equal to the sampling parameter times the population size. The sampling parameter is set to 0.06. The average initial sample size is 60.08, first wave is 272.50, and final sample is 332.58.

For the Bayes data augmentation estimators the burn-in is set to 10\% and 5000 samples are selected. MLEs based on full graph are: $\hat{\underline{\lambda}}_{MLE}=(4.800E-01, 5.200E-01)$ and $\hat{\underline{\beta}}_{MLE}=(1.193E-02, 9.696E-04, 6.025E-03)$. The tables below present the corresponding results from the simulation study. The prior for the population size is flat (prior 0).

\begin{table}[H]
\caption{Population size estimators for simulation study 2.3.}
\centering
\begin{tabular}{rrrrrrrr}
  \hline
Estimator            &Mean      &Var.     &MSE        &Coverage Rate    &Avg. Length           \\\hline
$\hat{N}_{LP}$       &1,002     &46,003   &46,008     &0.898            &787\\
$\hat{N}_1$          &878       &119,497  &134,378    &0.637            &733 \\
$\hat{N}_3$          &882       &11,293   &25,177     &0.645            &365 \\
$\hat{N}_5$          &883       &87,224   &100,849    &0.778            &1,047 \\
$\hat{N}_{Bayes,0}$  &931       &11,778   &16,499     &0.832            &388\\\hline
\label{2_3_Pop_size_ests}
\end{tabular}
\end{table}

\begin{table}[H]
\caption{Bayes data augmentation estimators for simulation study 2.3.}
\centering
\begin{tabular}{rrrrrrrr}
  \hline
Estimator            &Mean   &Var.                   &MSE         &Coverage Rate  &Avg. Length     \\\hline
$\hat{\lambda}_1$    & 5.368E-01 & 1.652E-03 & 4.881E-03 & 7.240E-01 & 1.596E-01 \\
$\hat{\lambda}_2$    & 4.632E-01 & 1.652E-03 & 4.881E-03 & 7.240E-01 & 1.596E-01 \\
$\hat{\beta}_{1,1}$  & 1.328E-02 & 2.726E-06 & 4.551E-06 & 8.730E-01 & 6.425E-03 \\
$\hat{\beta}_{1,2}$  & 1.350E-03 & 7.408E-08 & 2.188E-07 & 7.004E-01 & 1.074E-03 \\
$\hat{\beta}_{2,2}$  & 9.305E-03 & 3.330E-06 & 1.409E-05 & 4.532E-01 & 6.861E-03 \\ \hline
\label{2_3_prior_0_results}
\end{tabular}
\end{table}

\subsection{Simulation Study 3}

The network population is generated from a three-strata stochastic block model. Parameters are set to $(\lambda_1, \lambda_2, \lambda_3) = (0.5,0.4,0.1)$, $\beta_{11}=12/(N-2), \beta_{22}= 6/(N-2), \beta_{33}= 3/(N-2),  \beta_{12}=\beta_{13} = \beta_{23} = 1/(N-1)$.

\subsubsection{Study 3.1: Bernoulli Initial Sample, $\mathbf{N=100}$}

\begin{figure}[H]
	\centering
\vspace{-1mm}
\centering
		
		\includegraphics[scale=0.4]{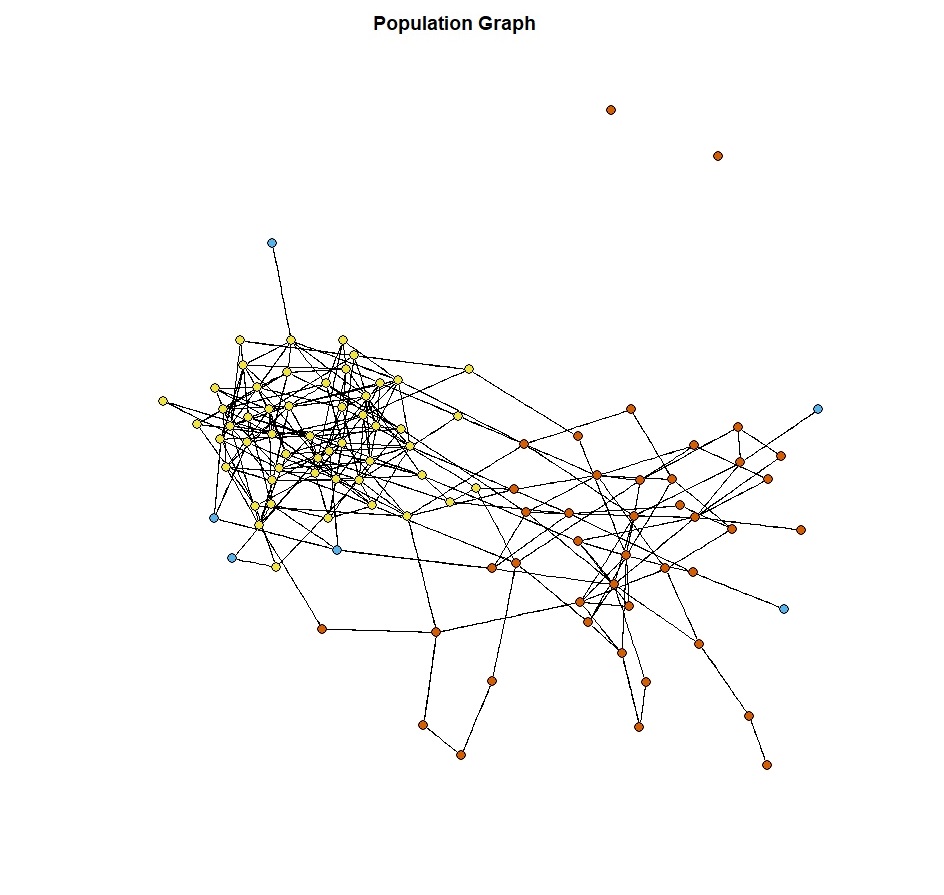}
\vspace{-3mm}
\caption{Population 3.1: Strata are represented by colour of nodes.}
\label{Pop3_1.jpeg}
\end{figure}

The sampling parameter is set to 0.20. The average initial sample size is 20.01, first wave is 49.40, and final sample is 69.41.

Gelman-Rubin statistics are based on chains of length 500, and 100 samples are selected to determine if the length of the chain is sufficient for estimating the population size and model parameters. The hyperparameter for the prior on the population size is set to zero. $\underline{\lambda}$ parameter seed values are set to (0.8,0.1,0.1) for the first chain and (0.1,0.1,0.8) for the second chain. $\underline{\beta}$ parameter seed values are set to 0.7 for the first chain and 0.3 for the second chain. The table below presents the mean and median of the corresponding Gelman-Rubin statistics.
\begin{table}[H]
\caption{Mean and median scores of Gelman-Rubin statistics corresponding with stochastic block model parameters for simulation study 3.1.}
\centering
\begin{tabular}{rrrrrrrr}
  \hline
Estimator            &Mean                  &Median                                   \\\hline
$\hat{N}$            &1.006                  &1.002                                \\
$\hat{\lambda}_1$    &1.004                  &1.002                                 \\
$\hat{\lambda}_2$    &1.004                  &1.002                                  \\
$\hat{\lambda}_3$    &1.003                  &1.001                                  \\
$\hat{\beta}_{1,1}$  &1.008                  &1.002   \\
$\hat{\beta}_{1,2}$  &1.003                  &1.001   \\
$\hat{\beta}_{1,3}$  &1.005                  &1.001   \\
$\hat{\beta}_{2,2}$  &1.008                  &1.004   \\
$\hat{\beta}_{2,3}$  &1.004                  &1.001   \\
$\hat{\beta}_{3,3}$  &1.002                  &1.005   \\\hline
\label{3_1_GR}
\end{tabular}
\end{table}

For the Bayes data augmentation estimators the burn-in is set to 10\% and 5000 samples are selected. MLEs based on full graph are: $\hat{\underline{\lambda}}_{MLE}=(5.200E-01, 4.200E-01, 6.000E-02)$ and
$\hat{\underline{\beta}}_{MLE}=(1.199E-01, 6.410E-03, 2.244E-02, 7.782E-02, 1.190E-02, 6.667E-02)$. The tables below present the corresponding results from the simulation study.

\begin{table}[H]
\caption{Population size estimators for simulation study 3.1.}
\centering
\begin{tabular}{rrrrrrrr}
  \hline
Estimator            &Mean      &Var.     &MSE        &Coverage Rate    &Avg. Length           \\\hline
$\hat{N}_{LP}$       &99.75     &449      &450        &0.903            &75.45\\
$\hat{N}_1$          &114.83    &3,572    &3,792      &0.877            &141.86\\
$\hat{N}_3$          &97.98     &178      &182        &0.911            &47.00\\
$\hat{N}_5$          &106.24    &1,653    &1,692      &0.930            &138.40 \\
$\hat{N}_{Bayes,0}$  &96.84     &122      &132        &0.899            &38.97 \\
$\hat{N}_{Bayes,1}$  &95.76     &110      &128        &0.891            &37.22 \\
$\hat{N}_{Bayes,2}$  &94.55     &101      &131        &0.873            &35.60 \\\hline
\label{3_1_Pop_size}
\end{tabular}
\end{table}

\begin{table}[H]
\caption{Bayes data augmentation estimators for simulation study 3.1. Population size is treated as known.}
\centering
\begin{tabular}{rrrrrrrr}
  \hline
Estimator            &Mean   &Var.     &MSE        &Coverage Rate  &Avg. Length     \\\hline
$\hat{\lambda}_1$    & 5.265E-01 & 1.789E-03 & 1.831E-03 & 9.962E-01 & 2.436E-01 \\
$\hat{\lambda}_2$    & 4.111E-01 & 2.158E-03 & 2.236E-03 & 9.926E-01 & 2.463E-01 \\
$\hat{\lambda}_3$    & 6.235E-02 & 7.065E-04 & 7.121E-04 & 9.616E-01 & 1.235E-01 \\
$\hat{\beta}_{1,1}$  & 1.199E-01 & 2.651E-04 & 2.651E-04 & 9.710E-01 & 7.029E-02 \\
$\hat{\beta}_{1,2}$  & 7.779E-03 & 7.138E-06 & 9.011E-06 & 9.694E-01 & 1.200E-02 \\
$\hat{\beta}_{1,3}$  & 6.099E-02 & 7.131E-03 & 8.617E-03 & 9.418E-01 & 1.336E-01 \\
$\hat{\beta}_{2,2}$  & 8.463E-02 & 4.517E-04 & 4.981E-04 & 9.322E-01 & 7.655E-02 \\
$\hat{\beta}_{2,3}$  & 5.048E-02 & 7.013E-03 & 8.501E-03 & 9.246E-01 & 1.287E-01 \\
$\hat{\beta}_{3,3}$  & 2.702E-01 & 2.099E-02 & 6.244E-02 & 9.818E-01 & 6.564E-01 \\\hline
\label{3_1_N_known_results}
\end{tabular}
\end{table}

\begin{table}[H]
\caption{Bayes data augmentation estimators for simulation study 3.1. Prior 0 used for prior distribution on population size.}
\centering
\begin{tabular}{rrrrrrrr}
  \hline
Estimator            &Mean   &Var.    &MSE       &Coverage Rate  &Avg. Length     \\\hline
$\hat{\lambda}_1$    & 5.373E-01 & 2.508E-03 & 2.806E-03 & 9.834E-01 & 2.540E-01 \\
$\hat{\lambda}_2$     & 4.025E-01 & 2.736E-03 & 3.043E-03 & 9.778E-01 & 2.521E-01 \\
$\hat{\lambda}_3$     & 6.025E-02 & 5.563E-04 & 5.564E-04 & 9.688E-01 & 1.198E-01 \\
$\hat{\beta}_{1,1}$   & 1.240E-01 & 3.536E-04 & 3.702E-04 & 9.696E-01 & 7.725E-02 \\
$\hat{\beta}_{1,2}$   & 8.346E-03 & 1.054E-05 & 1.429E-05 & 9.544E-01 & 1.345E-02 \\
$\hat{\beta}_{1,3}$   & 6.053E-02 & 6.056E-03 & 7.507E-03 & 9.446E-01 & 1.336E-01 \\
$\hat{\beta}_{2,2}$   & 9.464E-02 & 9.767E-04 & 1.260E-03 & 9.042E-01 & 9.839E-02 \\
$\hat{\beta}_{2,3}$   & 5.113E-02 & 6.134E-03 & 7.673E-03 & 9.230E-01 & 1.329E-01 \\
$\hat{\beta}_{3,3}$   & 2.803E-01 & 1.952E-02 & 6.517E-02 & 9.786E-01 & 6.753E-01 \\\hline
\label{3_1_prior_0_results}
\end{tabular}
\end{table}

\begin{table}[H]
\caption{Bayes data augmentation estimators for simulation study 3.1. Prior 1 used for prior distribution on population size.}
\centering
\begin{tabular}{rrrrrrrr}
  \hline
Estimator            &Mean   &Var.    &MSE       &Coverage Rate  &Avg. Length     \\\hline
$\hat{\lambda}_1$    & 5.395E-01 & 2.563E-03 & 2.942E-03 & 9.772E-01 & 2.520E-01 \\
$\hat{\lambda}_2$    & 4.002E-01 & 2.812E-03 & 3.205E-03 & 9.702E-01 & 2.500E-01 \\
$\hat{\lambda}_3$    & 6.036E-02 & 5.572E-04 & 5.573E-04 & 9.682E-01 & 1.189E-01 \\
$\hat{\beta}_{1,1}$  & 1.248E-01 & 3.578E-04 & 3.816E-04 & 9.678E-01 & 7.720E-02 \\
$\hat{\beta}_{1,2}$  & 8.486E-03 & 1.109E-05 & 1.540E-05 & 9.476E-01 & 1.364E-02 \\
$\hat{\beta}_{1,3}$  & 6.194E-02 & 6.444E-03 & 8.004E-03 & 9.356E-01 & 1.357E-01 \\
$\hat{\beta}_{2,2}$  & 9.577E-02 & 9.739E-04 & 1.296E-03 & 9.000E-01 & 9.861E-02 \\
$\hat{\beta}_{2,3}$  & 5.244E-02 & 6.401E-03 & 8.044E-03 & 9.136E-01 & 1.356E-01 \\
$\hat{\beta}_{3,3}$  & 2.806E-01 & 1.967E-02 & 6.543E-02 & 9.752E-01 & 6.759E-01 \\\hline
\label{3_1_prior_1_results}
\end{tabular}
\end{table}

\begin{table}[H]
\caption{Bayes data augmentation estimators for simulation study 3.1. Prior 2 used for prior distribution on population size.}
\centering
\begin{tabular}{rrrrrrrr}
  \hline
Estimator            &Mean   &Var.    &MSE       &Coverage Rate  &Avg. Length     \\\hline
$\hat{\lambda}_1$     & 5.406E-01 & 2.520E-03 & 2.946E-03 & 9.802E-01 & 2.501E-01 \\
$\hat{\lambda}_2$     & 3.995E-01 & 2.691E-03 & 3.109E-03 & 9.740E-01 & 2.479E-01 \\
$\hat{\lambda}_3$     & 5.981E-02 & 5.249E-04 & 5.250E-04 & 9.698E-01 & 1.178E-01 \\
$\hat{\beta}_{1,1}$   & 1.260E-01 & 3.570E-04 & 3.940E-04 & 9.680E-01 & 7.742E-02 \\
$\hat{\beta}_{1,2}$   & 8.615E-03 & 1.181E-05 & 1.667E-05 & 9.402E-01 & 1.379E-02 \\
$\hat{\beta}_{1,3}$   & 6.195E-02 & 6.388E-03 & 7.949E-03 & 9.404E-01 & 1.354E-01 \\
$\hat{\beta}_{2,2}$   & 9.705E-02 & 9.208E-04 & 1.291E-03 & 8.970E-01 & 9.877E-02 \\
$\hat{\beta}_{2,3}$   & 5.223E-02 & 6.229E-03 & 7.855E-03 & 9.156E-01 & 1.345E-01 \\
$\hat{\beta}_{3,3}$   & 2.827E-01 & 1.928E-02 & 6.596E-02 & 9.746E-01 & 6.793E-01 \\\hline
\label{3_1_prior_2_results}
\end{tabular}
\end{table}

\subsubsection{Study 3.2: Bernoulli Initial Sample, $\mathbf{N=1000}$}

\begin{figure}[H]
	\centering
\vspace{-1mm}
\centering
		
		\includegraphics[scale=0.4]{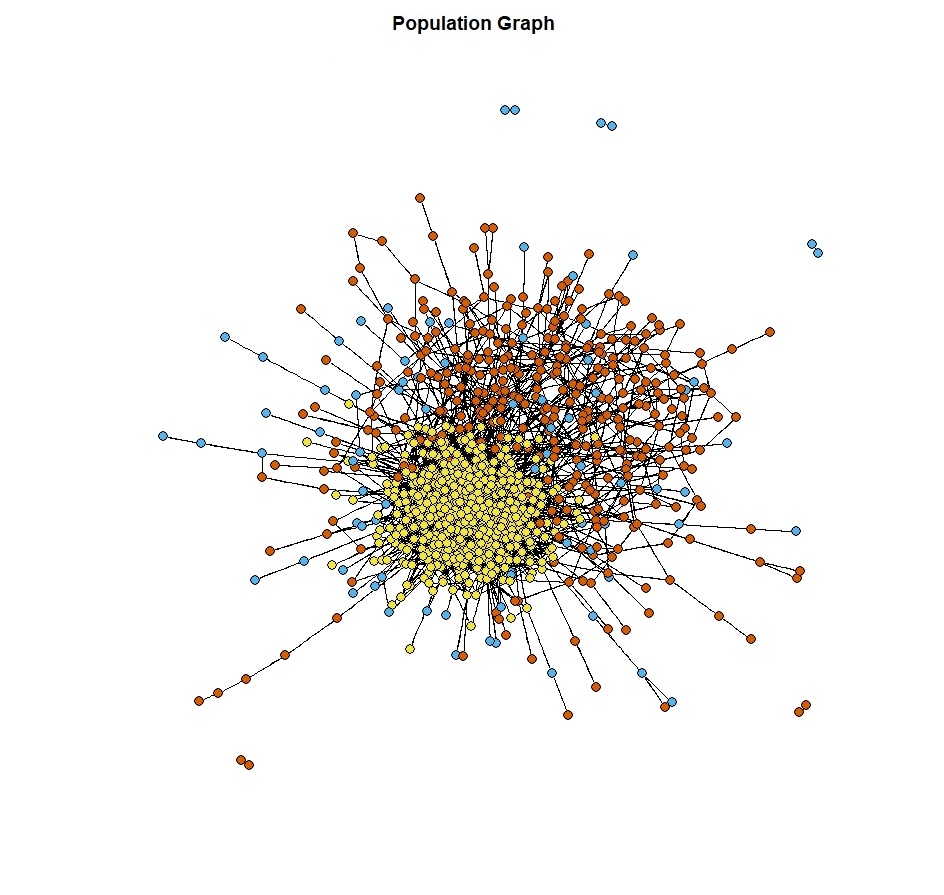}
\vspace{-3mm}
\caption{Population 3.2: Strata are represented by colour of nodes.}
\label{Pop3_2.jpeg}
\end{figure}

The sampling parameter is set to 0.06. The average initial sample size is 60.04, first wave is 225.14, and final sample is 285.18.

Gelman-Rubin statistics are based on chains of length 2000, and 100 samples are selected to determine if the length of the chain is sufficient for estimating the population size and model parameters. The hyperparameter for the prior on the population size is set to zero. $\underline{\lambda}$ parameter seed values are set to (0.8,0.1,0.1) for the first chain and (0.1,0.1,0.8) for the second chain. $\underline{\beta}$ parameter seed values are set to 0.7 for the first chain and 0.3 for the second chain. The table below presents the mean and median of the corresponding Gelman-Rubin statistics.
\begin{table}[H]
\caption{Mean and median scores of Gelman-Rubin statistics corresponding with stochastic block model parameters for simulation study 3.2.}
\centering
\begin{tabular}{rrrrrrrr}
  \hline
Estimator            &Mean                  &Median                                   \\\hline
$\hat{N}$            &1.016                  &1.007                                \\
$\hat{\lambda}_1$    &1.009                  &1.004                                 \\
$\hat{\lambda}_2$    &1.010                  &1.005                                  \\
$\hat{\lambda}_3$    &1.008                  &1.004                                  \\
$\hat{\beta}_{1,1}$  &1.016                  &1.005   \\
$\hat{\beta}_{1,2}$  &1.009                  &1.003   \\
$\hat{\beta}_{1,3}$  &1.006                  &1.003   \\
$\hat{\beta}_{2,2}$  &1.015                  &1.007   \\
$\hat{\beta}_{2,3}$  &1.006                  &1.002   \\
$\hat{\beta}_{3,3}$  &1.006                  &1.002   \\\hline
\label{3_2_GR}
\end{tabular}
\end{table}

For the Bayes data augmentation estimators the burn-in is set to 10\% and 5000 samples are selected. MLEs based on full graph are: $\hat{\underline{\lambda}}_{MLE}=(5.200E-01, 3.700E-01, 1.100E-01)$ and
$\hat{\underline{\beta}}_{MLE}=(1.218E-02, 9.927E-04, 8.916E-04, 5.816E-03, 1.057E-03, 3.670E-03).$ The tables below present the corresponding results from the simulation study.

\begin{table}[H]
\caption{Population size estimators for simulation study 3.2.}
\centering
\begin{tabular}{rrrrrrrr}
  \hline
Estimator            &Mean      &Var.     &MSE        &Coverage Rate    &Avg. Length           \\\hline
$\hat{N}_{LP}$       &1,004            &62,539   &62,559     &0.898            &897\\
$\hat{N}_1$          &1,193            &547,694  &584,838    &0.871            &1,654\\
$\hat{N}_3$          &903              &17,896   &27,429     &0.735            &460\\
$\hat{N}_5$          &1,150            &394,326  &416,678    &0.937            &2,171 \\
$\hat{N}_{Bayes,0}$  &940              &16,294   &19,902     &0.888            &487 \\
$\hat{N}_{Bayes,1}$  &923              &15,296   &21,295     &0.855            &468 \\
$\hat{N}_{Bayes,2}$  &905              &14,403   &23,392     &0.822            &451 \\\hline
\label{3_2_Pop_size}
\end{tabular}
\end{table}

\begin{table}[H]
\caption{Bayes data augmentation estimators for simulation study 3.2. Population size is treated as known.}
\centering
\begin{tabular}{rrrrrrrr}
  \hline
Estimator            &Mean   &Var.     &MSE        &Coverage Rate  &Avg. Length     \\\hline
$\hat{\lambda}_1$    & 5.261E-01 & 1.723E-03 & 1.761E-03 & 9.492E-01 & 1.653E-01 \\
$\hat{\lambda}_2$    & 3.691E-01 & 1.828E-03 & 1.829E-03 & 9.494E-01 & 1.674E-01 \\
$\hat{\lambda}_3$    & 1.048E-01 & 9.890E-04 & 1.017E-03 & 9.162E-01 & 1.157E-01 \\
$\hat{\beta}_{1,1}$  & 1.224E-02 & 1.679E-06 & 1.682E-06 & 9.482E-01 & 5.081E-03 \\
$\hat{\beta}_{1,2}$  & 1.043E-03 & 4.839E-08 & 5.087E-08 & 9.444E-01 & 8.565E-04 \\
$\hat{\beta}_{1,3}$  & 1.179E-03 & 3.373E-07 & 4.199E-07 & 9.540E-01 & 1.875E-03 \\
$\hat{\beta}_{2,2}$  & 6.145E-03 & 1.432E-06 & 1.541E-06 & 9.442E-01 & 4.553E-03 \\
$\hat{\beta}_{2,3}$  & 1.447E-03 & 7.073E-07 & 8.600E-07 & 9.164E-01 & 2.433E-03 \\
$\hat{\beta}_{3,3}$  & 7.836E-03 & 2.635E-04 & 2.808E-04 & 9.416E-01 & 2.019E-02 \\\hline
\label{3_2_N_known_results}
\end{tabular}
\end{table}

\begin{table}[H]
\caption{Bayes data augmentation estimators for simulation study 3.2. Prior 0 used for prior distribution on population size.}
\centering
\begin{tabular}{rrrrrrrr}
  \hline
Estimator            &Mean   &Var.                   &MSE            &Coverage Rate  &Avg. Length     \\\hline
$\hat{\lambda}_1$    & 5.392E-01 & 2.148E-03 & 2.518E-03 & 9.282E-01 & 1.777E-01 \\
$\hat{\lambda}_2$    & 3.601E-01 & 1.929E-03 & 2.028E-03 & 9.388E-01 & 1.703E-01 \\
$\hat{\lambda}_3$    & 1.007E-01 & 9.649E-04 & 1.052E-03 & 9.064E-01 & 1.134E-01 \\
$\hat{\beta}_{1,1}$  & 1.307E-02 & 3.419E-06 & 4.208E-06 & 9.248E-01 & 7.098E-03 \\
$\hat{\beta}_{1,2}$  & 1.159E-03 & 8.472E-08 & 1.123E-07 & 9.242E-01 & 1.121E-03 \\
$\hat{\beta}_{1,3}$  & 1.320E-03 & 4.390E-07 & 6.224E-07 & 9.368E-01 & 2.225E-03 \\
$\hat{\beta}_{2,2}$  & 7.042E-03 & 3.197E-06 & 4.700E-06 & 9.136E-01 & 6.909E-03 \\
$\hat{\beta}_{2,3}$  & 1.703E-03 & 1.153E-06 & 1.572E-06 & 9.038E-01 & 3.151E-03 \\
$\hat{\beta}_{3,3}$  & 9.596E-03 & 4.294E-04 & 4.646E-04 & 9.276E-01 & 2.576E-02 \\\hline
\label{3_2_prior_0_results}
\end{tabular}
\end{table}

\begin{table}[H]
\caption{Bayes data augmentation estimators for simulation study 3.2. Prior 1 used for prior distribution on population size.}
\centering
\begin{tabular}{rrrrrrrr}
  \hline
Estimator            &Mean   &Var.    &MSE       &Coverage Rate  &Avg. Length     \\\hline
$\hat{\lambda}_1$     & 5.415E-01 & 2.063E-03 & 2.523E-03 & 9.184E-01 & 1.762E-01 \\
$\hat{\lambda}_2$     & 3.581E-01 & 1.889E-03 & 2.030E-03 & 9.384E-01 & 1.687E-01 \\
$\hat{\lambda}_3$     & 1.004E-01 & 8.854E-04 & 9.776E-04 & 9.160E-01 & 1.127E-01 \\
$\hat{\beta}_{1,1}$   & 1.328E-02 & 3.436E-06 & 4.635E-06 & 9.138E-01 & 7.139E-03 \\
$\hat{\beta}_{1,2}$   & 1.180E-03 & 8.833E-08 & 1.233E-07 & 9.098E-01 & 1.135E-03 \\
$\hat{\beta}_{1,3}$   & 1.323E-03 & 4.445E-07 & 6.307E-07 & 9.346E-01 & 2.220E-03 \\
$\hat{\beta}_{2,2}$   & 7.213E-03 & 3.521E-06 & 5.473E-06 & 8.992E-01 & 7.026E-03 \\
$\hat{\beta}_{2,3}$   & 1.715E-03 & 1.093E-06 & 1.527E-06 & 8.980E-01 & 3.161E-03 \\
$\hat{\beta}_{3,3}$   & 9.537E-03 & 3.124E-04 & 3.468E-04 & 9.180E-01 & 2.531E-02 \\\hline
\label{3_2_prior_1_results}
\end{tabular}
\end{table}

\begin{table}[H]
\caption{Bayes data augmentation estimators for simulation study 3.2. Prior 2 used for prior distribution on population size.}
\centering
\begin{tabular}{rrrrrrrr}
  \hline
Estimator            &Mean   &Var.               &MSE                &Coverage Rate  &Avg. Length     \\\hline
$\hat{\lambda}_1$   & 5.436E-01 & 2.091E-03 & 2.648E-03 & 9.160E-01 & 1.750E-01 \\
$\hat{\lambda}_2$   & 3.567E-01 & 1.941E-03 & 2.118E-03 & 9.308E-01 & 1.677E-01 \\
$\hat{\lambda}_3$   & 9.970E-02 & 9.224E-04 & 1.028E-03 & 9.026E-01 & 1.113E-01 \\
$\hat{\beta}_{1,1}$ & 1.346E-02 & 3.712E-06 & 5.346E-06 & 9.016E-01 & 7.172E-03 \\
$\hat{\beta}_{1,2}$ & 1.197E-03 & 9.473E-08 & 1.365E-07 & 9.026E-01 & 1.152E-03 \\
$\hat{\beta}_{1,3}$ & 1.349E-03 & 3.480E-07 & 5.570E-07 & 9.246E-01 & 2.249E-03 \\
$\hat{\beta}_{2,2}$ & 7.374E-03 & 3.684E-06 & 6.112E-06 & 8.722E-01 & 7.123E-03 \\
$\hat{\beta}_{2,3}$ & 1.750E-03 & 9.327E-07 & 1.414E-06 & 8.846E-01 & 3.207E-03 \\
$\hat{\beta}_{3,3}$ & 9.493E-03 & 2.369E-04 & 2.708E-04 & 9.128E-01 & 2.545E-02 \\\hline
\label{3_2_prior_2_results}
\end{tabular}
\end{table}

\subsubsection{Study 3.3: Unequal Probability Initial Sample, $\mathbf{N=1000}$}
\begin{figure}[H]
	\centering
\vspace{-1mm}
\centering
		
		\includegraphics[scale=0.4]{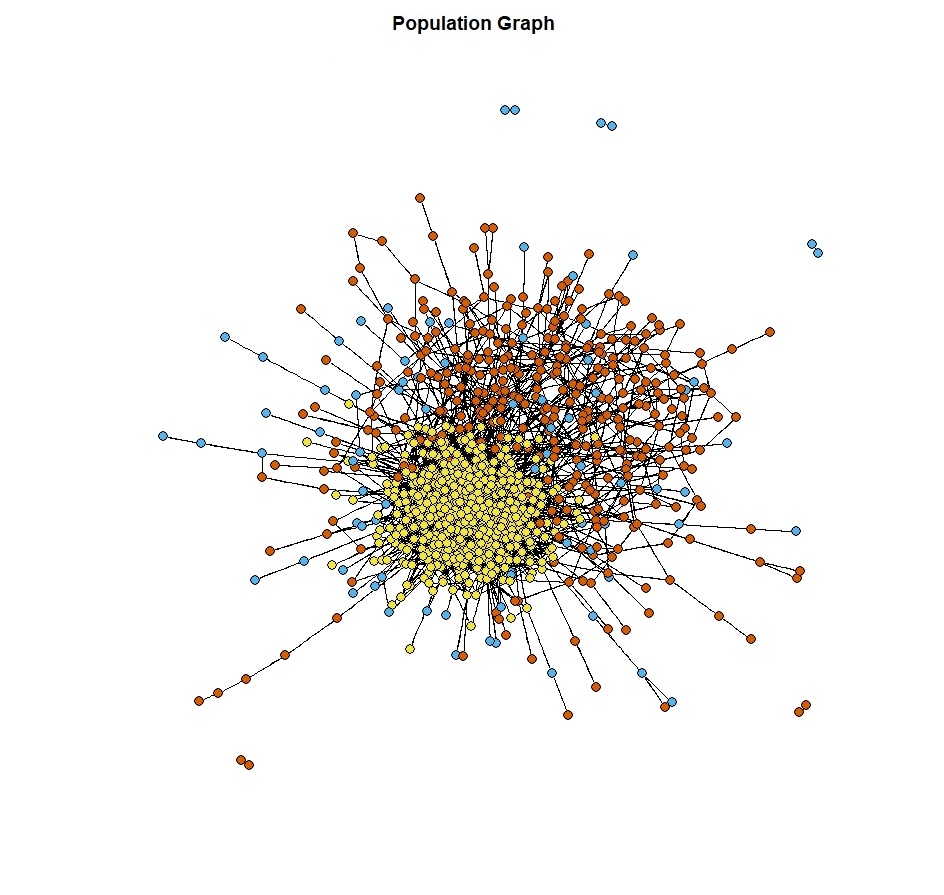}
\vspace{-3mm}
\caption{Population 3.3: Strata are represented by colour of nodes.}
\label{Pop3_2.jpeg}
\end{figure}

Selection for the initial sample is with probability proportional to individual degree plus one where probabilities are scaled so the expected size of the initial sample is equal to the sampling parameter times the population size. The sampling parameter is set to 0.06. The average initial sample size is 60.23, first wave is 275.78, and final sample is 336.01.

For the Bayes data augmentation estimators the burn-in is set to 10\% and 5000 samples are selected. MLEs based on full graph are: $\hat{\underline{\lambda}}_{MLE}=(5.200E-01, 3.700E-01, 1.100E-01)$ and
$\hat{\underline{\beta}}_{MLE}=(1.218E-02, 9.927E-04, 8.916E-04, 5.816E-03, 1.057E-03, 3.670E-03)$. The tables below present the corresponding results from the simulation study. The prior for the population size is flat (prior 0).

\begin{table}[H]
\caption{Population size estimators for simulation study 3.3.}
\centering
\begin{tabular}{rrrrrrrr}
  \hline
Estimator            &Mean      &Var.     &MSE        &Coverage Rate    &Avg. Length           \\\hline
$\hat{N}_{LP}$       &1,002     &45,245   &45,250     &0.930            &782\\
$\hat{N}_1$          &776       &71,358   &121,560    &0.453            &578 \\
$\hat{N}_3$          &761       &6,321    &63,206     &0.151            &276 \\
$\hat{N}_5$          &785       &50,131   &96,213     &0.590            &785 \\
$\hat{N}_{Bayes,0}$  &815       &7,041    &41,445     &0.398            &301 \\\hline
\label{3_3_Pop_size_ests}
\end{tabular}
\end{table}

\begin{table}[H]
\caption{Bayes data augmentation estimators for simulation study 3.3.}
\centering
\begin{tabular}{rrrrrrrr}
  \hline
Estimator            &Mean   &Var.                   &MSE         &Coverage Rate  &Avg. Length     \\\hline
$\hat{\lambda}_1$     & 6.437E-01 & 1.835E-03 & 1.713E-02 & 1.820E-01 & 1.601E-01 \\
$\hat{\lambda}_2$     & 2.990E-01 & 1.776E-03 & 6.819E-03 & 5.836E-01 & 1.528E-01 \\
$\hat{\lambda}_3$     & 5.734E-02 & 5.081E-04 & 3.281E-03 & 4.178E-01 & 7.778E-02 \\
$\hat{\beta}_{1,1}$   & 1.371E-02 & 1.847E-06 & 4.172E-06 & 8.124E-01 & 5.529E-03 \\
$\hat{\beta}_{1,2}$   & 1.686E-03 & 1.650E-07 & 6.456E-07 & 4.642E-01 & 1.480E-03 \\
$\hat{\beta}_{1,3}$   & 3.161E-03 & 7.352E-06 & 1.250E-05 & 4.968E-01 & 5.888E-03 \\
$\hat{\beta}_{2,2}$   & 1.242E-02 & 1.121E-05 & 5.488E-05 & 1.978E-01 & 1.204E-02 \\
$\hat{\beta}_{2,3}$   & 5.170E-03 & 2.619E-05 & 4.311E-05 & 4.902E-01 & 1.134E-02 \\
$\hat{\beta}_{3,3}$   & 6.250E-02 & 1.492E-02 & 1.838E-02 & 5.906E-01 & 1.505E-01 \\\hline
\label{3_3_prior_0_results}
\end{tabular}
\end{table}

\subsection{Simulation Study 4}

The network population is generated from a four-strata stochastic block model. Parameters are set to $(\lambda_1, \lambda_2, \lambda_3, \lambda_4) = (0.3,0.3,0.3,0.1)$, $\beta_{11}=12/(N-2), \beta_{22}= 9/(N-2), \beta_{33}= 6/(N-2), \beta_{44}=3/(N-2), \beta_{12}=\beta_{13}=\beta_{23}=\beta_{24}=\beta_{34}=1/(N-1)$.

\subsubsection{Study 4.1: Bernoulli Initial Sample, $\mathbf{N=100}$}

\begin{figure}[H]
	\centering
\vspace{-1mm}
\centering
		
		\includegraphics[scale=0.4]{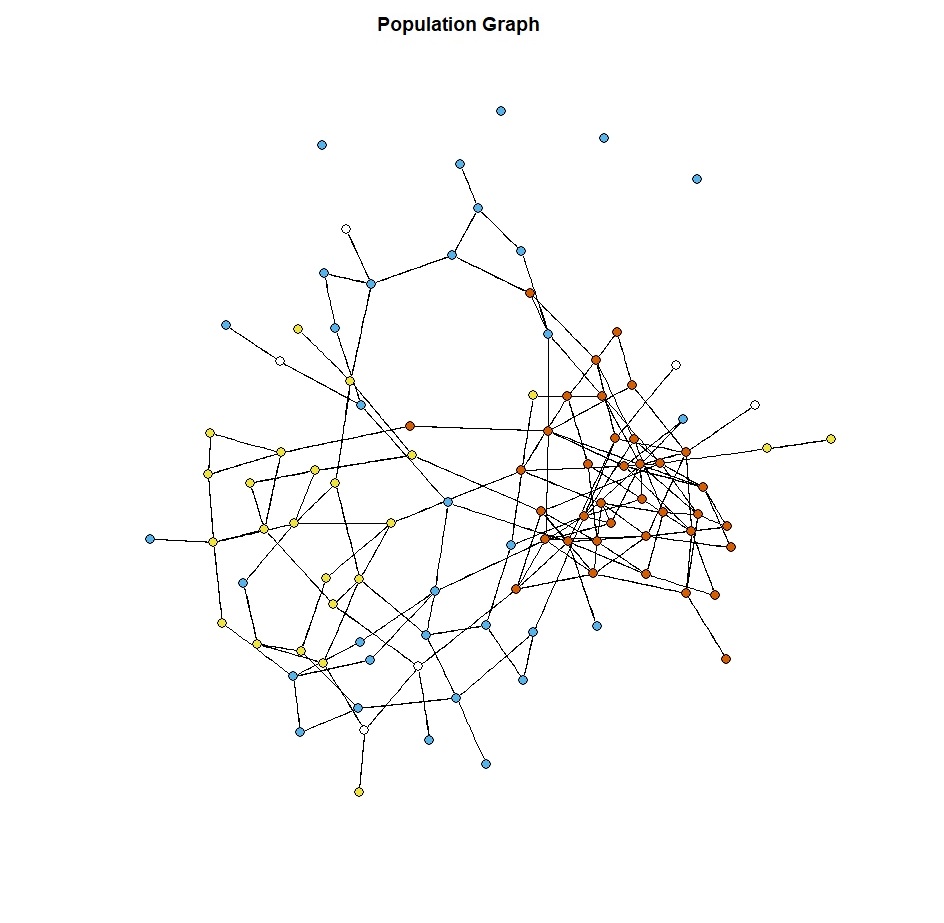}
\vspace{-3mm}
\caption{Population 4.1: Strata are represented by colour of nodes.}
\label{Pop4_1.jpeg}
\end{figure}

The sampling parameter is set to 0.20. The average initial sample size is 20.01, first wave is 38.30, and final sample is 58.31.

Gelman-Rubin statistics are based on chains of length 500, and 100 samples are selected to determine if the length of the chain is sufficient for estimating the population size and model parameters. The hyperparameter for the prior on the population size is set to zero. $\underline{\lambda}$ parameter seed values are set to (0.7,0.1,0.1,0.1) for the first chain and (0.1,0.1,0.1,0.7) for the second chain. $\underline{\beta}$ parameter seed values are set to 0.7 for the first chain and 0.3 for the second chain. The table below presents the mean and median of the corresponding Gelman-Rubin statistics.
\begin{table}[H]
\caption{Mean and median scores of Gelman-Rubin statistics corresponding with stochastic block model parameters for simulation study 4.1.}
\centering
\begin{tabular}{rrrrrrrr}
  \hline
Estimator            &Mean                  &Median                                   \\\hline
$\hat{N}$            &1.007                  &1.002   \\
$\hat{\lambda}_1$    &1.002                  &1.000    \\
$\hat{\lambda}_2$    &1.003                  &1.000     \\
$\hat{\lambda}_3$    &1.003                  &1.000     \\
$\hat{\lambda}_4$    &1.002                  &1.000     \\
$\hat{\beta}_{1,1}$  &1.006                  &1.003   \\
$\hat{\beta}_{1,2}$  &1.005                  &1.001   \\
$\hat{\beta}_{1,3}$  &1.004                  &1.001   \\
$\hat{\beta}_{1,4}$  &1.004                  &1.000   \\
$\hat{\beta}_{2,2}$  &1.008                  &1.003   \\
$\hat{\beta}_{2,3}$  &1.003                  &1.001   \\
$\hat{\beta}_{2,4}$  &1.004                  &1.000   \\
$\hat{\beta}_{3,3}$  &1.007                  &1.003   \\
$\hat{\beta}_{3,4}$  &1.003                  &1.001   \\
$\hat{\beta}_{4,4}$  &1.001                  &1.000     \\\hline
\label{4_1_GR}
\end{tabular}
\end{table}

For the Bayes data augmentation estimators the burn-in is set to 10\% and 5000 samples are selected. MLEs based on full graph are: $\hat{\underline{\lambda}}_{MLE}=(2.400E-01, 3.700E-01, 3.330E-01, 6.000E-02)$ and
$\hat{\underline{\beta}}_{MLE}=(9.783E-02, 6.757E-03, 8.838E-03, 2.083E-02, 1.141E-01, 9.828E-03, 1.351E-02, 4.735E-02, 2.020E-02, 6.667E-02)$. The tables below present the corresponding results from the simulation study.

\begin{table}[H]
\caption{Population size estimators for simulation study 4.1.}
\centering
\begin{tabular}{rrrrrrrr}
  \hline
Estimator            &Mean      &Var.     &MSE        &Coverage Rate    &Avg. Length           \\\hline
$\hat{N}_{LP}$       &99.45            &300      &300        &0.926            &62.59\\
$\hat{N}_1$          &121.39           &5,609    &6,067      &0.877            &234.94\\
$\hat{N}_3$          &103.42           &475      &487        &0.927            &71.41\\
$\hat{N}_5$          &112.54           &3,123    &3,280      &0.923            &206.06 \\
$\hat{N}_{Bayes,0}$  &88.68            &172      &300        &0.806            &46.00 \\
$\hat{N}_{Bayes,1}$  &87.20            &158      &322        &0.766            &43.39 \\
$\hat{N}_{Bayes,2}$  &85.72            &156      &360        &0.721            &40.92 \\
\hline
\label{4_1_Pop_size}
\end{tabular}
\end{table}

\begin{table}[H]
\caption{Bayes data augmentation estimators for simulation study 4.1. Population size is treated as known.}
\centering
\begin{tabular}{rrrrrrrr}
  \hline
Estimator            &Mean   &Var.               &MSE                &Coverage Rate  &Avg. Length     \\\hline
$\hat{\lambda}_1$     & 2.404E-01 & 3.161E-03 & 3.161E-03 & 9.660E-01 & 2.396E-01 \\
$\hat{\lambda}_2$     & 3.727E-01 & 3.124E-03 & 3.131E-03 & 9.786E-01 & 2.560E-01 \\
$\hat{\lambda}_3$     & 3.294E-01 & 4.050E-03 & 4.051E-03 & 9.712E-01 & 2.692E-01 \\
$\hat{\lambda}_4$     & 5.741E-02 & 6.141E-04 & 6.208E-04 & 9.366E-01 & 1.150E-01 \\
$\hat{\beta}_{1,1}$   & 1.229E-01 & 3.447E-03 & 4.076E-03 & 9.698E-01 & 1.843E-01 \\
$\hat{\beta}_{1,2}$   & 1.104E-02 & 1.011E-04 & 1.194E-04 & 9.590E-01 & 2.428E-02 \\
$\hat{\beta}_{1,3}$   & 1.388E-02 & 1.210E-04 & 1.464E-04 & 9.774E-01 & 2.901E-02 \\
$\hat{\beta}_{1,4}$   & 8.014E-02 & 8.744E-03 & 1.226E-02 & 8.872E-01 & 1.979E-01 \\
$\hat{\beta}_{2,2}$   & 1.214E-01 & 9.577E-04 & 1.011E-03 & 9.564E-01 & 1.095E-01 \\
$\hat{\beta}_{2,3}$   & 1.262E-02 & 2.816E-05 & 3.593E-05 & 9.512E-01 & 2.130E-02 \\
$\hat{\beta}_{2,4}$   & 5.556E-02 & 6.990E-03 & 8.759E-03 & 9.130E-01 & 1.398E-01 \\
$\hat{\beta}_{3,3}$   & 5.726E-02 & 6.238E-04 & 7.220E-04 & 9.576E-01 & 8.594E-02 \\
$\hat{\beta}_{3,4}$   & 6.784E-02 & 7.414E-03 & 9.684E-03 & 9.152E-01 & 1.626E-01 \\
$\hat{\beta}_{4,4}$   & 2.771E-01 & 1.956E-02 & 6.383E-02 & 9.784E-01 & 6.709E-01 \\\hline
\label{4_1_N_known_results}
\end{tabular}
\end{table}

\begin{table}[H]
\caption{Bayes data augmentation estimators for simulation study 4.1. Prior 0 used for prior distribution on population size.}
\centering
\begin{tabular}{rrrrrrrr}
  \hline
Estimator            &Mean   &Var.               &MSE                &Coverage Rate  &Avg. Length     \\\hline
$\hat{\lambda}_1$   & 2.388E-01 & 3.142E-03 & 3.143E-03 & 9.534E-01 & 2.289E-01 \\
$\hat{\lambda}_2$   & 3.904E-01 & 3.737E-03 & 4.151E-03 & 9.640E-01 & 2.581E-01 \\
$\hat{\lambda}_3$   & 3.130E-01 & 3.799E-03 & 4.088E-03 & 9.532E-01 & 2.587E-01 \\
$\hat{\lambda}_4$   & 5.782E-02 & 5.429E-04 & 5.476E-04 & 9.550E-01 & 1.125E-01 \\
$\hat{\beta}_{1,1}$ & 1.451E-01 & 5.268E-03 & 7.503E-03 & 9.372E-01 & 2.139E-01 \\
$\hat{\beta}_{1,2}$ & 1.318E-02 & 1.640E-04 & 2.052E-04 & 9.388E-01 & 2.931E-02 \\
$\hat{\beta}_{1,3}$ & 1.750E-02 & 1.919E-04 & 2.669E-04 & 9.476E-01 & 3.875E-02 \\
$\hat{\beta}_{1,4}$ & 9.073E-02 & 9.738E-03 & 1.462E-02 & 8.598E-01 & 2.235E-01 \\
$\hat{\beta}_{2,2}$  & 1.342E-01 & 1.426E-03 & 1.828E-03 & 9.402E-01 & 1.226E-01 \\
$\hat{\beta}_{2,3}$  & 1.536E-02 & 5.699E-05 & 8.758E-05 & 9.148E-01 & 2.723E-02 \\
$\hat{\beta}_{2,4}$  & 6.266E-02 & 7.828E-03 & 1.024E-02 & 8.900E-01 & 1.564E-01 \\
$\hat{\beta}_{3,3}$  & 7.309E-02 & 1.435E-03 & 2.098E-03 & 9.256E-01 & 1.159E-01 \\
$\hat{\beta}_{3,4}$  & 7.872E-02 & 8.314E-03 & 1.174E-02 & 8.774E-01 & 1.893E-01 \\
$\hat{\beta}_{4,4}$  & 2.946E-01 & 1.845E-02 & 7.040E-02 & 9.670E-01 & 6.951E-01 \\\hline
\label{4_1_prior_0_results}
\end{tabular}
\end{table}

\begin{table}[H]
\caption{Bayes data augmentation estimators for simulation study 4.1. Prior 1 used for prior distribution on population size.}
\centering
\begin{tabular}{rrrrrrrr}
  \hline
Estimator            &Mean   &Var.               &MSE                &Coverage Rate  &Avg. Length     \\\hline
$\hat{\lambda}_1$   & 2.388E-01 & 3.033E-03 & 3.035E-03 & 9.568E-01 & 2.267E-01 \\
$\hat{\lambda}_2$   & 3.921E-01 & 3.655E-03 & 4.144E-03 & 9.574E-01 & 2.563E-01 \\
$\hat{\lambda}_3$   & 3.114E-01 & 3.576E-03 & 3.923E-03 & 9.562E-01 & 2.560E-01 \\
$\hat{\lambda}_4$   & 5.770E-02 & 5.242E-04 & 5.295E-04 & 9.580E-01 & 1.121E-01 \\
$\hat{\beta}_{1,1}$ & 1.475E-01 & 5.050E-03 & 7.520E-03 & 9.288E-01 & 2.142E-01 \\
$\hat{\beta}_{1,2}$ & 1.307E-02 & 1.725E-04 & 2.124E-04 & 9.406E-01 & 2.899E-02 \\
$\hat{\beta}_{1,3}$ & 1.774E-02 & 2.201E-04 & 2.994E-04 & 9.412E-01 & 3.884E-02 \\
$\hat{\beta}_{1,4}$ & 9.164E-02 & 9.922E-03 & 1.493E-02 & 8.600E-01 & 2.247E-01 \\
$\hat{\beta}_{2,2}$  & 1.353E-01 & 1.388E-03 & 1.835E-03 & 9.398E-01 & 1.223E-01 \\
$\hat{\beta}_{2,3}$  & 1.554E-02 & 5.905E-05 & 9.166E-05 & 9.136E-01 & 2.746E-02 \\
$\hat{\beta}_{2,4}$  & 6.388E-02 & 8.060E-03 & 1.060E-02 & 8.896E-01 & 1.583E-01 \\
$\hat{\beta}_{3,3}$  & 7.490E-02 & 1.433E-03 & 2.192E-03 & 9.124E-01 & 1.166E-01 \\
$\hat{\beta}_{3,4}$  & 7.974E-02 & 8.315E-03 & 1.186E-02 & 8.822E-01 & 1.912E-01 \\
$\hat{\beta}_{4,4}$  & 2.969E-01 & 1.846E-02 & 7.148E-02 & 9.688E-01 & 6.976E-01 \\ \hline
  \label{4_1_prior_1_results}
\end{tabular}
\end{table}

\begin{table}[H]
\caption{Bayes data augmentation estimators for simulation study 4.1. Prior 2 used for prior distribution on population size.}
\centering
\begin{tabular}{rrrrrrrr}
  \hline
Estimator            &Mean   &Var.               &MSE                &Coverage Rate  &Avg. Length     \\\hline
$\hat{\lambda}_1$    & 2.380E-01 & 2.933E-03 & 2.937E-03 & 9.550E-01 & 2.248E-01 \\
$\hat{\lambda}_2$    & 3.936E-01 & 3.678E-03 & 4.236E-03 & 9.600E-01 & 2.550E-01 \\
$\hat{\lambda}_3$    & 3.102E-01 & 3.509E-03 & 3.902E-03 & 9.550E-01 & 2.535E-01 \\
$\hat{\lambda}_4$    & 5.820E-02 & 5.119E-04 & 5.152E-04 & 9.612E-01 & 1.126E-01 \\
$\hat{\beta}_{1,1}$  & 1.505E-01 & 5.513E-03 & 8.291E-03 & 9.232E-01 & 2.159E-01 \\
$\hat{\beta}_{1,2}$  & 1.335E-02 & 1.313E-04 & 1.748E-04 & 9.272E-01 & 2.950E-02 \\
$\hat{\beta}_{1,3}$  & 1.790E-02 & 1.766E-04 & 2.587E-04 & 9.366E-01 & 3.926E-02 \\
$\hat{\beta}_{1,4}$  & 8.934E-02 & 9.147E-03 & 1.384E-02 & 8.614E-01 & 2.199E-01 \\
$\hat{\beta}_{2,2}$   & 1.376E-01 & 1.524E-03 & 2.077E-03 & 9.306E-01 & 1.225E-01 \\
$\hat{\beta}_{2,3}$   & 1.579E-02 & 5.618E-05 & 9.176E-05 & 9.078E-01 & 2.767E-02 \\
$\hat{\beta}_{2,4}$   & 6.184E-02 & 7.671E-03 & 1.001E-02 & 8.940E-01 & 1.539E-01 \\
$\hat{\beta}_{3,3}$   & 7.607E-02 & 1.540E-03 & 2.365E-03 & 9.068E-01 & 1.171E-01 \\
$\hat{\beta}_{3,4}$   & 7.862E-02 & 7.959E-03 & 1.137E-02 & 8.740E-01 & 1.875E-01 \\
$\hat{\beta}_{4,4}$   & 2.953E-01 & 1.793E-02 & 7.019E-02 & 9.668E-01 & 6.957E-01 \\ \hline
  \label{4_1_prior_2_results}
\end{tabular}
\end{table}

\subsubsection{Study 4.2: Bernoulli Initial Sample, $\mathbf{N=1000}$}

\begin{figure}[H]
	\centering
\vspace{-1mm}
\centering
		
		\includegraphics[scale=0.4]{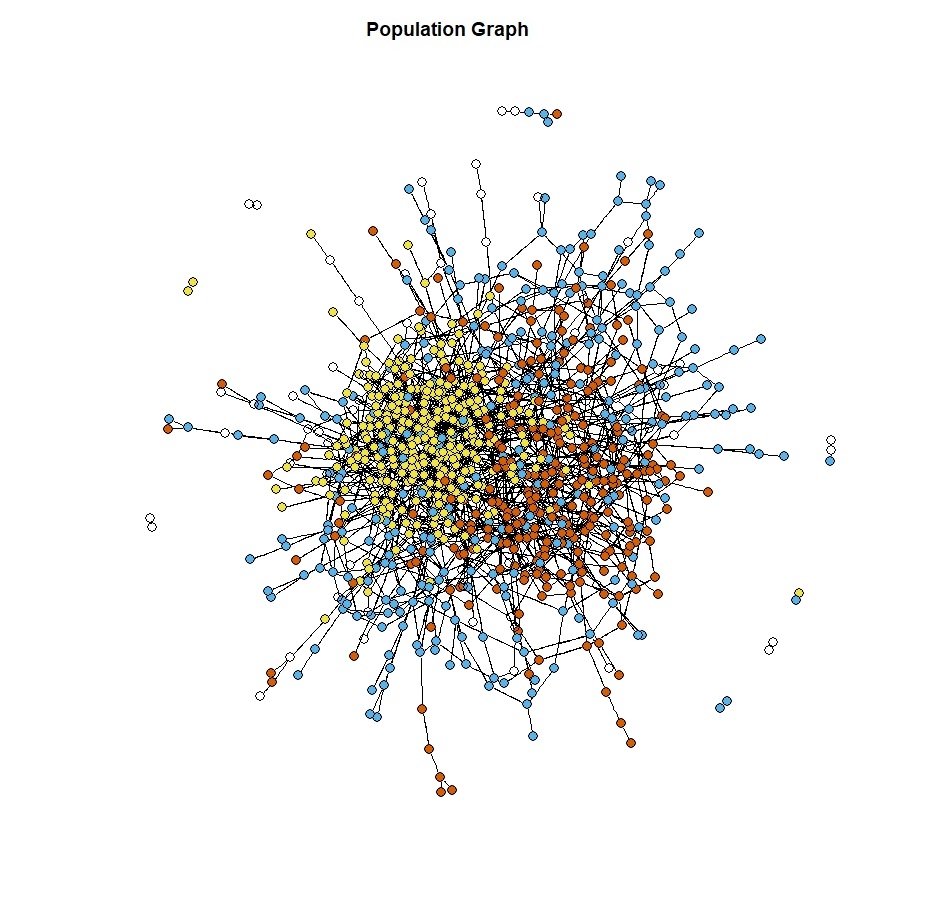}
\vspace{-3mm}
\caption{Population 4.2: Strata are represented by colour of nodes.}
\label{Pop4_2.jpeg}
\end{figure}

The sampling parameter is set to 0.06. The average initial sample size is 60.04, first wave is 152.79, and final sample is 212.83.

Gelman-Rubin statistics are based on chains of length 2000, and 100 samples are selected to determine if the length of the chain is sufficient for estimating the population size and model parameters. The hyperparameter for the prior on the population size is set to zero. $\underline{\lambda}$ parameter seed values are set to (0.7,0.1,0.1,0.1) for the first chain and (0.1,0.1,0.1,0.7) for the second chain. $\underline{\beta}$ parameter seed values are set to 0.7 for the first chain and 0.3 for the second chain. The table below presents the mean and median of the corresponding Gelman-Rubin statistics.
\begin{table}[H]
\caption{Mean and median scores of Gelman-Rubin statistics corresponding with stochastic block model parameters for simulation study 4.2.}
\centering
\begin{tabular}{rrrrrrrr}
  \hline
Estimator            &Mean                  &Median                                   \\\hline
$\hat{N}$            &1.014                  &1.008                                \\
$\hat{\lambda}_1$    &1.007                  &1.003                                 \\
$\hat{\lambda}_2$    &1.007                  &1.003                              \\
$\hat{\lambda}_3$    &1.006                  &1.002                                  \\
$\hat{\lambda}_4$    &1.005                  &1.003                                  \\
$\hat{\beta}_{1,1}$  &1.010                  &1.005   \\
$\hat{\beta}_{1,2}$  &1.005                  &1.002   \\
$\hat{\beta}_{1,3}$  &1.004                  &1.001   \\
$\hat{\beta}_{1,4}$  &1.003                  &1.001   \\
$\hat{\beta}_{2,2}$  &1.013                  &1.007   \\
$\hat{\beta}_{2,3}$  &1.006                  &1.003   \\
$\hat{\beta}_{2,4}$  &1.004                  &1.002   \\
$\hat{\beta}_{3,3}$  &1.009                  &1.004   \\
$\hat{\beta}_{3,4}$  &1.004                  &1.002   \\
$\hat{\beta}_{4,4}$  &1.005                  &1.002       \\\hline
\label{4_2_GR}
\end{tabular}
\end{table}

For the Bayes data augmentation estimators the burn-in is set to 10\% and 5000 samples are selected. MLEs based on full graph are: $\hat{\underline{\lambda}}_{MLE}=(2.930E-01, 2.840E-01, 3.130E-01, 1.100E-01)$
and $\hat{\underline{\beta}}_{MLE}=(1.213E-02, 9.614E-04, 8.832E-04, 8.377E-04, 8.162E-03, 9.787E-04, 9.603E-04, 5.652E-03, 1.075E-03, 3.670E-03)$. The tables below present the corresponding results from the simulation study.

\begin{table}[H]
\caption{Population size estimators for simulation study 4.2.}
\centering
\begin{tabular}{rrrrrrrr}
  \hline
Estimator            &Mean      &Var.           &MSE            &Coverage Rate      &Avg. Length           \\\hline
$\hat{N}_{LP}$       &1,005     &116,449        &116,471        &0.890              &1,132\\
$\hat{N}_1$          &1,320     &1,135,717      &1,237,879      &0.863              &4,574\\
$\hat{N}_3$          &1,009     &67,911         &67,989         &0.873              &797\\
$\hat{N}_5$          &1,275     &929,083        &1,004,958      &0.924              &3,294 \\
$\hat{N}_{Bayes,0}$   &768       &17,687         &71,652         &0.566              &502 \\
$\hat{N}_{Bayes,1}$  &746       &16,207         &80,938         &0.500              &475 \\
$\hat{N}_{Bayes,2}$  &722       &14,272         &91,474         &0.421              &452 \\
\hline
\label{4_2_Pop_size}
\end{tabular}
\end{table}

\begin{table}[H]
\caption{Bayes data augmentation estimators for simulation study 4.2. Population size is treated as known.}
\centering
\begin{tabular}{rrrrrr}
  \hline
Estimator            &Mean   &Var.               &MSE                &Coverage Rate  &Avg. Length     \\\hline
  \hline
$\hat{\lambda}_1$     & 2.948E-01 & 1.994E-03 & 1.998E-03 & 9.348E-01 & 1.647E-01 \\
$\hat{\lambda}_2$     & 2.841E-01 & 2.063E-03 & 2.063E-03 & 9.340E-01 & 1.688E-01 \\
$\hat{\lambda}_3$     & 3.176E-01 & 2.278E-03 & 2.300E-03 & 9.344E-01 & 1.780E-01 \\
$\hat{\lambda}_4$     & 1.034E-01 & 1.105E-03 & 1.149E-03 & 9.014E-01 & 1.169E-01 \\
$\hat{\beta}_{1,1}$   & 1.283E-02 & 7.142E-06 & 7.633E-06 & 9.376E-01 & 9.764E-03 \\
$\hat{\beta}_{1,2}$   & 1.104E-03 & 1.272E-07 & 1.474E-07 & 9.370E-01 & 1.380E-03 \\
$\hat{\beta}_{1,3}$   & 9.965E-04 & 9.409E-08 & 1.069E-07 & 9.544E-01 & 1.235E-03 \\
$\hat{\beta}_{1,4}$   & 1.295E-03 & 8.958E-07 & 1.105E-06 & 9.416E-01 & 2.604E-03 \\
$\hat{\beta}_{2,2}$    & 8.847E-03 & 4.850E-06 & 5.318E-06 & 9.328E-01 & 7.865E-03 \\
$\hat{\beta}_{2,3}$    & 1.108E-03 & 1.165E-07 & 1.332E-07 & 9.456E-01 & 1.332E-03 \\
$\hat{\beta}_{2,4}$    & 1.477E-03 & 1.152E-06 & 1.420E-06 & 9.050E-01 & 2.847E-03 \\
$\hat{\beta}_{3,3}$    & 6.011E-03 & 2.012E-06 & 2.140E-06 & 9.574E-01 & 5.462E-03 \\
$\hat{\beta}_{3,4}$    & 1.522E-03 & 9.382E-07 & 1.138E-06 & 9.364E-01 & 2.756E-03 \\
$\hat{\beta}_{4,4}$    & 8.350E-03 & 2.796E-04 & 3.015E-04 & 9.452E-01 & 2.217E-02 \\
   \hline
\end{tabular}
\end{table}

\begin{table}[H]
\caption{Bayes data augmentation estimators for simulation study 4.2. Prior 0 used for prior distribution on population size.}
\centering
\begin{tabular}{rrrrrr}
  \hline
Estimator            &Mean   &Var.               &MSE                &Coverage Rate  &Avg. Length     \\\hline
  \hline
$\hat{\lambda}_1$    & 3.077E-01 & 2.021E-03 & 2.236E-03 & 9.172E-01 & 1.605E-01 \\
$\hat{\lambda}_2$    & 2.846E-01 & 1.888E-03 & 1.889E-03 & 9.348E-01 & 1.595E-01 \\
$\hat{\lambda}_3$    & 3.101E-01 & 2.068E-03 & 2.077E-03 & 9.272E-01 & 1.673E-01 \\
$\hat{\lambda}_4$    & 9.757E-02 & 9.310E-04 & 1.086E-03 & 8.872E-01 & 1.087E-01 \\
$\hat{\beta}_{1,1}$  & 1.686E-02 & 1.689E-05 & 3.922E-05 & 7.734E-01 & 1.503E-02 \\
$\hat{\beta}_{1,2}$  & 1.489E-03 & 3.023E-07 & 5.808E-07 & 8.378E-01 & 2.071E-03 \\
$\hat{\beta}_{1,3}$  & 1.375E-03 & 2.344E-07 & 4.766E-07 & 8.550E-01 & 1.915E-03 \\
$\hat{\beta}_{1,4}$  & 1.802E-03 & 1.036E-06 & 1.966E-06 & 8.622E-01 & 3.795E-03 \\
$\hat{\beta}_{2,2}$   & 1.217E-02 & 1.272E-05 & 2.880E-05 & 7.810E-01 & 1.297E-02 \\
$\hat{\beta}_{2,3}$   & 1.562E-03 & 3.107E-07 & 6.505E-07 & 8.320E-01 & 2.171E-03 \\
$\hat{\beta}_{2,4}$   & 2.080E-03 & 1.626E-06 & 2.880E-06 & 8.304E-01 & 4.291E-03 \\
$\hat{\beta}_{3,3}$   & 8.626E-03 & 6.409E-06 & 1.525E-05 & 7.910E-01 & 9.717E-03 \\
$\hat{\beta}_{3,4}$   & 2.240E-03 & 1.705E-06 & 3.062E-06 & 8.364E-01 & 4.390E-03 \\
$\hat{\beta}_{4,4}$   & 1.189E-02 & 3.489E-04 & 4.166E-04 & 8.668E-01 & 3.197E-02 \\
   \hline
\end{tabular}
\end{table}

\begin{table}[H]
\caption{Bayes data augmentation estimators for simulation study 4.2. Prior 1 used for prior distribution on population size.}
\centering
\begin{tabular}{rrrrrr}
  \hline
Estimator            &Mean   &Var.               &MSE                &Coverage Rate  &Avg. Length     \\\hline
  \hline
$\hat{\lambda}_1$    & 3.096E-01 & 2.036E-03 & 2.311E-03 & 9.078E-01 & 1.601E-01 \\
$\hat{\lambda}_2$    & 2.851E-01 & 1.978E-03 & 1.979E-03 & 9.278E-01 & 1.588E-01 \\
$\hat{\lambda}_3$    & 3.084E-01 & 2.096E-03 & 2.117E-03 & 9.274E-01 & 1.662E-01 \\
$\hat{\lambda}_4$    & 9.686E-02 & 9.044E-04 & 1.077E-03 & 8.850E-01 & 1.076E-01 \\
$\hat{\beta}_{1,1}$  & 1.718E-02 & 1.696E-05 & 4.241E-05 & 7.524E-01 & 1.514E-02 \\
$\hat{\beta}_{1,2}$  & 1.524E-03 & 3.050E-07 & 6.214E-07 & 8.268E-01 & 2.109E-03 \\
$\hat{\beta}_{1,3}$  & 1.404E-03 & 2.393E-07 & 5.104E-07 & 8.332E-01 & 1.954E-03 \\
$\hat{\beta}_{1,4}$  & 1.865E-03 & 1.352E-06 & 2.407E-06 & 8.456E-01 & 3.914E-03 \\
$\hat{\beta}_{2,2}$   & 1.250E-02 & 1.290E-05 & 3.167E-05 & 7.446E-01 & 1.317E-02 \\
$\hat{\beta}_{2,3}$   & 1.608E-03 & 3.344E-07 & 7.308E-07 & 8.094E-01 & 2.223E-03 \\
$\hat{\beta}_{2,4}$   & 2.157E-03 & 2.005E-06 & 3.436E-06 & 8.148E-01 & 4.444E-03 \\
$\hat{\beta}_{3,3}$   & 8.922E-03 & 6.970E-06 & 1.766E-05 & 7.524E-01 & 9.958E-03 \\
$\hat{\beta}_{3,4}$   & 2.311E-03 & 1.677E-06 & 3.207E-06 & 8.310E-01 & 4.523E-03 \\
$\hat{\beta}_{4,4}$   & 1.244E-02 & 4.175E-04 & 4.944E-04 & 8.592E-01 & 3.342E-02 \\
   \hline
\end{tabular}
\end{table}

\begin{table}[H]
\caption{Bayes data augmentation estimators for simulation study 4.2. Prior 2 used for prior distribution on population size.}
\centering
\begin{tabular}{rrrrrr}
  \hline
Estimator            &Mean   &Var.               &MSE                &Coverage Rate  &Avg. Length     \\\hline
  \hline
$\hat{\lambda}_1$   & 3.113E-01 & 1.978E-03 & 2.314E-03 & 9.082E-01 & 1.596E-01 \\
$\hat{\lambda}_2$   & 2.851E-01 & 1.926E-03 & 1.927E-03 & 9.214E-01 & 1.577E-01 \\
$\hat{\lambda}_3$   & 3.077E-01 & 2.049E-03 & 2.077E-03 & 9.328E-01 & 1.652E-01 \\
$\hat{\lambda}_4$   & 9.592E-02 & 9.045E-04 & 1.103E-03 & 8.748E-01 & 1.065E-01 \\
$\hat{\beta}_{1,1}$ & 1.762E-02 & 1.770E-05 & 4.776E-05 & 7.112E-01 & 1.541E-02 \\
$\hat{\beta}_{1,2}$ & 1.565E-03 & 3.068E-07 & 6.711E-07 & 8.168E-01 & 2.159E-03 \\
$\hat{\beta}_{1,3}$ & 1.453E-03 & 2.592E-07 & 5.844E-07 & 8.116E-01 & 2.010E-03 \\
$\hat{\beta}_{1,4}$ & 1.909E-03 & 1.440E-06 & 2.589E-06 & 8.436E-01 & 4.022E-03 \\
$\hat{\beta}_{2,2}$  & 1.291E-02 & 1.324E-05 & 3.578E-05 & 7.092E-01 & 1.345E-02 \\
$\hat{\beta}_{2,3}$  & 1.654E-03 & 3.509E-07 & 8.065E-07 & 7.870E-01 & 2.276E-03 \\
$\hat{\beta}_{2,4}$  & 2.250E-03 & 2.213E-06 & 3.876E-06 & 8.000E-01 & 4.596E-03 \\
$\hat{\beta}_{3,3}$  & 9.184E-03 & 7.005E-06 & 1.948E-05 & 7.200E-01 & 1.019E-02 \\
$\hat{\beta}_{3,4}$  & 2.395E-03 & 1.954E-06 & 3.696E-06 & 8.072E-01 & 4.672E-03 \\
$\hat{\beta}_{4,4}$  & 1.301E-02 & 4.268E-04 & 5.141E-04 & 8.418E-01 & 3.491E-02 \\
   \hline
\end{tabular}
\end{table}

\subsubsection{Study 4.3: Unequal Probability Initial Sample, $\mathbf{N=1000}$}
\begin{figure}[H]
	\centering
\vspace{-1mm}
\centering
		
		\includegraphics[scale=0.4]{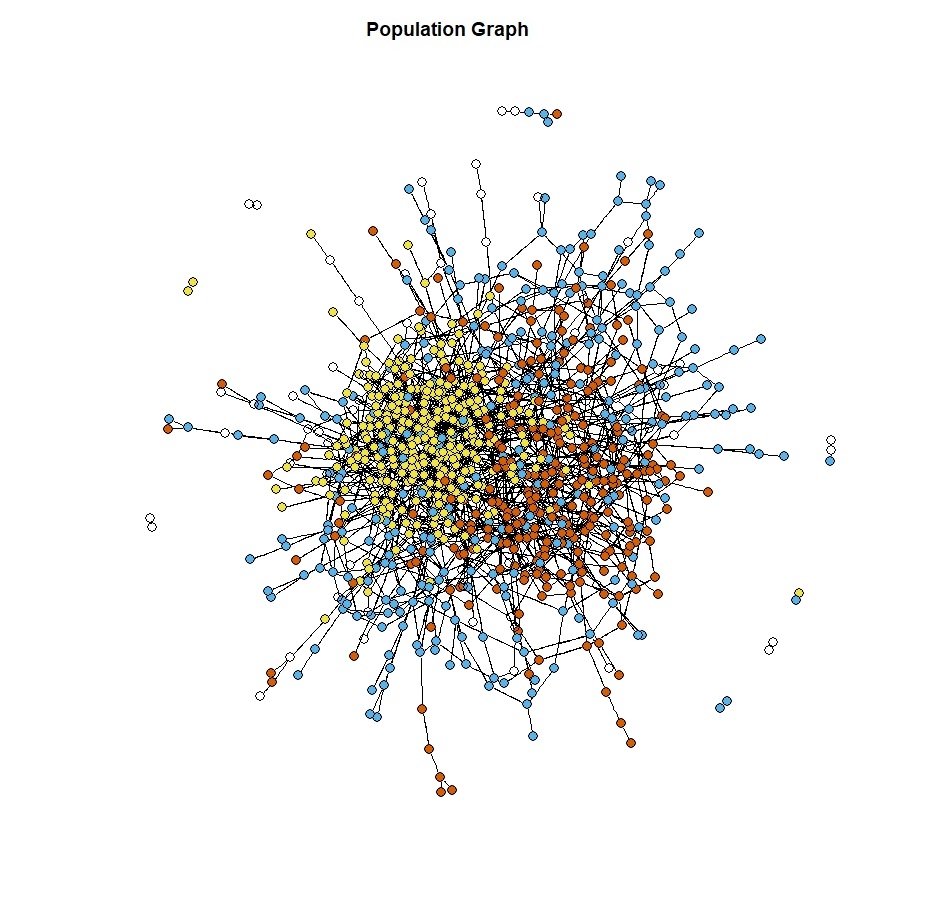}
\vspace{-3mm}
\caption{Population 4.3: Strata are represented by colour of nodes.}
\label{Pop4_2.jpeg}
\end{figure}

Selection for the initial sample is with probability proportional to individual degree plus one where probabilities are scaled so the expected size of the initial sample is equal to the sampling parameter times the population size. The sampling parameter is set to 0.06. The average initial sample size is 60.12, first wave is 190.49, and final sample is 250.61.

For the Bayes data augmentation estimators the burn-in is set to 10\% and 5000 samples are selected. MLEs based on full graph are: $\hat{\underline{\lambda}}_{MLE}=(2.930E-01, 2.840E-01, 3.130E-01, 1.100E-01)$
and $\hat{\underline{\beta}}_{MLE}=(1.213E-02, 9.614E-04, 8.832E-04, 8.377E-04, 8.162E-03, 9.787E-04, 9.603E-04, 5.652E-03, 1.075E-03, 3.670E-03)$. The tables below present the corresponding results from the simulation study. The prior for the population size is flat (prior 0).

\begin{table}[H]
\caption{Population size estimators for simulation study 4.3.}
\centering
\begin{tabular}{rrrrrrrr}
  \hline
Estimator            &Mean      &Var.     &MSE        &Coverage Rate    &Avg. Length           \\\hline
$\hat{N}_{LP}$       &1,004     &80,207   &80,224     &0.917            &995\\
$\hat{N}_1$          &855       &203,658  &224,608    &0.593            &1,017 \\
$\hat{N}_3$          &854       &23,350   &44,453     &0.626            &488 \\
$\hat{N}_5$          &856       &161,287  &182,098    &0.742            &1,414 \\
$\hat{N}_{Bayes,0}$  &762       &11,245   &67,655     &0.424            &399\\\hline
\label{4_3_Pop_size_ests}
\end{tabular}
\end{table}

\begin{table}[H]
\caption{Bayes data augmentation estimators for simulation study 4.3.}
\centering
\begin{tabular}{rrrrrrrr}
  \hline
Estimator            &Mean   &Var.                   &MSE         &Coverage Rate  &Avg. Length     \\\hline
$\hat{\lambda}_1$      & 3.570E-01 & 1.740E-03 & 5.834E-03 & 6.316E-01 & 1.560E-01 \\
$\hat{\lambda}_2$    & 2.908E-01 & 1.594E-03 & 1.640E-03 & 9.402E-01 & 1.502E-01 \\
$\hat{\lambda}_3$    & 2.835E-01 & 1.769E-03 & 2.638E-03 & 8.670E-01 & 1.542E-01 \\
$\hat{\lambda}_4$    & 6.871E-02 & 5.854E-04 & 2.290E-03 & 5.830E-01 & 8.404E-02 \\
$\hat{\beta}_{1,1}$  & 1.642E-02 & 8.852E-06 & 2.721E-05 & 6.906E-01 & 1.169E-02 \\
$\hat{\beta}_{1,2}$  & 1.697E-03 & 2.399E-07 & 7.805E-07 & 6.352E-01 & 1.954E-03 \\
$\hat{\beta}_{1,3}$  & 1.643E-03 & 2.361E-07 & 8.134E-07 & 6.244E-01 & 1.979E-03 \\
$\hat{\beta}_{1,4}$  & 2.837E-03 & 5.979E-06 & 9.975E-06 & 6.464E-01 & 5.814E-03 \\
$\hat{\beta}_{2,2}$   & 1.449E-02 & 1.262E-05 & 5.260E-05 & 4.240E-01 & 1.344E-02 \\
$\hat{\beta}_{2,3}$   & 1.977E-03 & 4.173E-07 & 1.413E-06 & 5.770E-01 & 2.503E-03 \\
$\hat{\beta}_{2,4}$   & 3.602E-03 & 7.982E-06 & 1.496E-05 & 5.778E-01 & 7.422E-03 \\
$\hat{\beta}_{3,3}$   & 1.146E-02 & 9.499E-06 & 4.318E-05 & 3.364E-01 & 1.204E-02 \\
$\hat{\beta}_{3,4}$   & 4.293E-03 & 1.200E-05 & 2.236E-05 & 5.080E-01 & 8.548E-03 \\
$\hat{\beta}_{4,4}$   & 3.322E-02 & 4.672E-03 & 5.545E-03 & 6.180E-01 & 8.549E-02 \\ \hline
\label{4_3_prior_0_results}
\end{tabular}
\end{table}

\subsection{Simulation Study 5}

The network population is generated from a three-strata stochastic block model. The population size is set to $N=1000$. Parameters are set to $(\lambda_1, \lambda_2, \lambda_3) = (0.3,0.3,0.4)$, $\beta_{11}=12/(N-2), \beta_{22}= 9/(N-2), \beta_{33}= 6/(N-2), \beta_{12}=\beta_{13}=\beta_{23}=1/(N-1)$.

\subsubsection{Study 5.1: Bernoulli Initial Sample, $\mathbf{N=1000}$}

\begin{figure}[H]
	\centering
\vspace{-1mm}
\centering
		
		\includegraphics[scale=0.4]{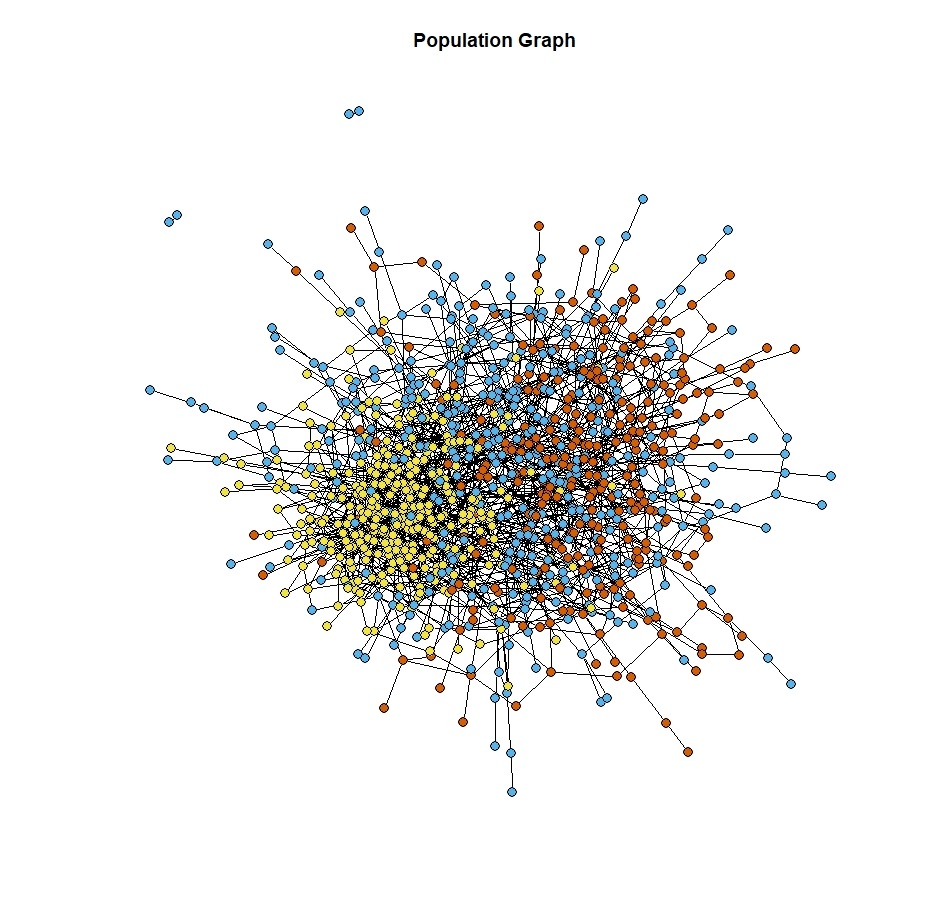}
\vspace{-3mm}
\caption{Population 5.1: Strata are represented by colour of nodes.}
\label{Pop5.jpeg}
\end{figure}

The sampling parameter is set to 0.025. The average initial sample size is 25.06, first wave is 80.44, second wave is 214.62 and final sample is 320.11.

Gelman-Rubin statistics are based on chains of length 2000, and 100 samples are selected to determine if the length of the chain is sufficient for estimating the population size and model parameters. The hyperparameter for the prior on the population size is set to zero. $\underline{\lambda}$ parameter seed values are set to (0.8,0.1,0.1) for the first chain and (0.1,0.1,0.8) for the second chain. $\underline{\beta}$ parameter seed values are set to 0.7 for the first chain and 0.3 for the second chain. The table below presents the mean and median of the corresponding Gelman-Rubin statistics.
\begin{table}[H]
\caption{Mean and median scores of Gelman-Rubin statistics corresponding with stochastic block model parameters for simulation study 5.1.}
\centering
\begin{tabular}{rrrrrrrr}
  \hline
Estimator            &Mean                   &Median                                   \\\hline
$\hat{N}$            &1.034                  &1.011                                \\
$\hat{\lambda}_1$    &1.014                  &1.004                                 \\
$\hat{\lambda}_2$    &1.014                  &1.006                              \\
$\hat{\lambda}_3$    &1.014                  &1.004                                 \\
$\hat{\beta}_{1,1}$  &1.028                  &1.010   \\
$\hat{\beta}_{1,2}$  &1.017                  &1.007   \\
$\hat{\beta}_{1,3}$  &1.014                  &1.005   \\
$\hat{\beta}_{2,2}$  &1.027                  &1.008   \\
$\hat{\beta}_{2,3}$  &1.020                  &1.004   \\
$\hat{\beta}_{3,3}$  &1.023                  &1.006   \\\hline
\end{tabular}
\end{table}

MLEs based on full graph are: $\hat{\underline{\lambda}}_{MLE}=(3.040E-01, 2.910E-01, 4.050E-01)$ and
$\hat{\underline{\beta}}_{MLE}=(1.205E-02, 9.721E-04, 9.747E-04, 8.698E-03, 9.334E-04, 5.892E-03)$. The tables below present the corresponding results from the simulation study. The prior for the population size is flat (prior 0).

\begin{table}[H]
\caption{Population size estimators for simulation study 5.1.}
\centering
\begin{tabular}{rrrrrrrr}
  \hline
Estimator                           &Mean     &Var.       &MSE        &Coverage Rate    &Avg. Length           \\\hline
$\hat{N}_{LP}$                      &882             &370,335    &384,169    &0.674            &1,861\\
$\hat{N}_1$                         &1,186           &528,727    &563,487    &0.843           &14,235\\
$\hat{N}_3$                         &1,568           &3,157,742  &3,480,428  &0.890            &$\infty$ \\
$\hat{N}_5$                         &1,146           &467,637    &488,809    &0.608            &1,963 \\
$\hat{N}_{Bayes,\ 1\ wave}$           &585             &42,154     &214,723    &0.403            &670 \\
$\hat{N}_{Bayes,\ 2\ waves}$          &939             &15,760     &19,434     &0.882            &453 \\\hline
\label{5_1_Pop_size_ests}
\end{tabular}
\end{table}

\begin{table}[H]
\caption{Bayes data augmentation estimators for simulation study 5.1 based on one wave of sampling.}
\centering
\begin{tabular}{rrrrrr}
  \hline
Estimator            &Mean   &Var.               &MSE                &Coverage Rate  &Avg. Length     \\\hline
  \hline
$\hat{\lambda}_1$    & 3.129E-01 & 6.108E-03 & 6.187E-03 & 8.866E-01 & 2.441E-01 \\
$\hat{\lambda}_2$    & 2.865E-01 & 5.932E-03 & 5.952E-03 & 8.860E-01 & 2.424E-01 \\
$\hat{\lambda}_3$    & 4.006E-01 & 6.532E-03 & 6.551E-03 & 9.058E-01 & 2.651E-01 \\
$\hat{\beta}_{1,1}$  & 2.953E-02 & 5.289E-04 & 8.345E-04 & 5.796E-01 & 4.614E-02 \\
$\hat{\beta}_{1,2}$  & 2.903E-03 & 1.077E-05 & 1.450E-05 & 7.290E-01 & 6.292E-03 \\
$\hat{\beta}_{1,3}$  & 2.653E-03 & 2.931E-06 & 5.747E-06 & 6.796E-01 & 5.154E-03 \\
$\hat{\beta}_{2,2}$  & 2.365E-02 & 4.473E-04 & 6.709E-04 & 6.134E-01 & 4.245E-02 \\
$\hat{\beta}_{2,3}$  & 2.642E-03 & 3.117E-06 & 6.035E-06 & 6.992E-01 & 5.350E-03 \\
$\hat{\beta}_{3,3}$  & 1.448E-02 & 6.318E-05 & 1.370E-04 & 5.698E-01 & 2.323E-02 \\
   \hline
\end{tabular}
\end{table}

\begin{table}[H]
\caption{Bayes data augmentation estimators for simulation study 5.1 based on two waves of sampling.}
\centering
\begin{tabular}{rrrrrr}
  \hline
Estimator            &Mean   &Var.               &MSE                &Coverage Rate  &Avg. Length     \\\hline
  \hline
$\hat{\lambda}_1$      & 3.141E-01 & 1.605E-03 & 1.708E-03 & 9.378E-01 & 1.497E-01 \\
$\hat{\lambda}_2$      & 2.864E-01 & 1.968E-03 & 1.989E-03 & 9.244E-01 & 1.571E-01 \\
$\hat{\lambda}_3$      & 3.994E-01 & 2.077E-03 & 2.108E-03 & 9.474E-01 & 1.724E-01 \\
$\hat{\beta}_{1,1}$    & 1.311E-02 & 6.542E-06 & 7.657E-06 & 9.398E-01 & 8.809E-03 \\
$\hat{\beta}_{1,2}$    & 1.198E-03 & 1.301E-07 & 1.811E-07 & 9.116E-01 & 1.275E-03 \\
$\hat{\beta}_{1,3}$    & 1.144E-03 & 8.107E-08 & 1.098E-07 & 9.320E-01 & 1.092E-03 \\
$\hat{\beta}_{2,2}$    & 1.021E-02 & 7.223E-06 & 9.520E-06 & 9.272E-01 & 9.333E-03 \\
$\hat{\beta}_{2,3}$    & 1.200E-03 & 1.121E-07 & 1.831E-07 & 8.932E-01 & 1.248E-03 \\
$\hat{\beta}_{3,3}$    & 6.901E-03 & 2.329E-06 & 3.347E-06 & 9.196E-01 & 5.753E-03 \\
   \hline
\end{tabular}
\end{table}

\subsubsection{Study 5.2: Unequal Probability Initial Sample, $\mathbf{N=1000}$}
\begin{figure}[H]
	\centering
\vspace{-1mm}
\centering
		
		\includegraphics[scale=0.4]{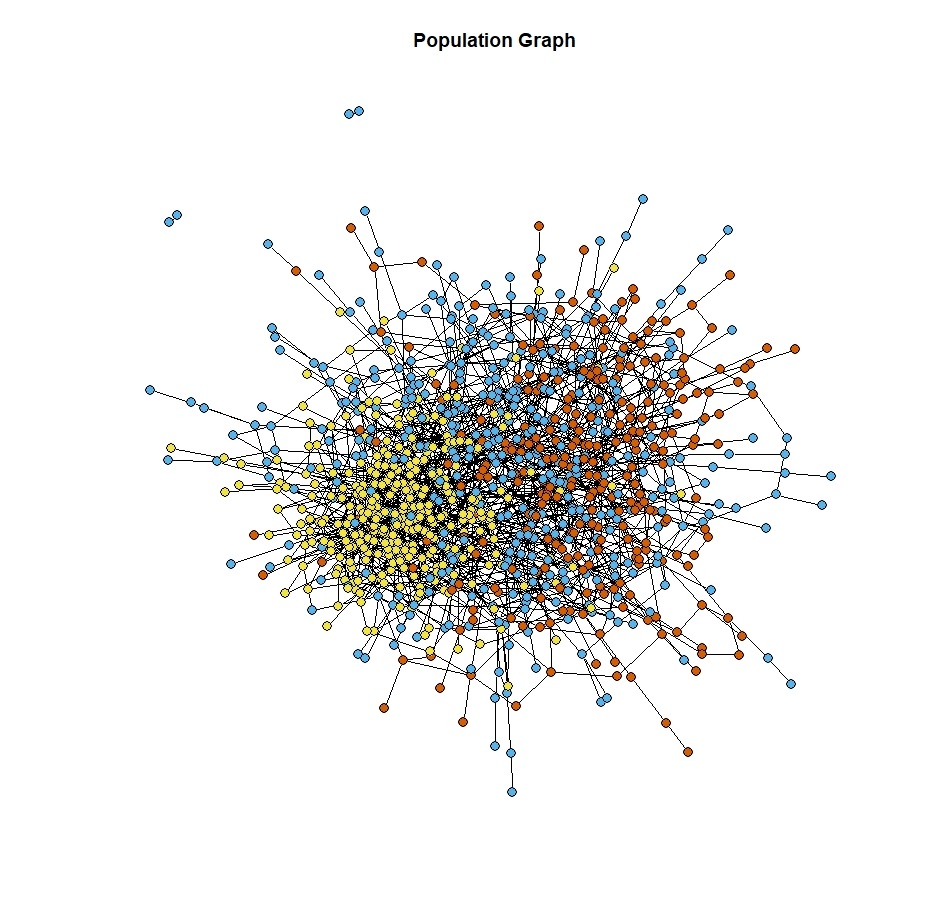}
\vspace{-3mm}
\caption{Population 5.2: Strata are represented by colour of nodes.}
\label{Pop5_2}
\end{figure}

Selection for the initial sample is with probability proportional to individual degree plus one where probabilities are scaled so the expected size of the initial sample is equal to the sampling parameter times the population size. The sampling parameter is set to 0.025. The average initial sample size is 25.09, first wave is 98.62, second wave is 245.23, and final sample is 368.95.

For the Bayes data augmentation estimators the burn-in is set to 10\% and 5000 samples are selected. MLEs based on full graph are: $\hat{\underline{\lambda}}_{MLE}=(3.040E-01, 2.910E-01, 4.050E-01)$ and
$\hat{\underline{\beta}}_{MLE}=(1.205E-02, 9.721E-04, 9.747E-04, 8.698E-03, 9.334E-04, 5.892E-03)$. The tables below present the corresponding results from the simulation study. The prior distribution for the population size is flat (prior 0).

\begin{table}[H]
\caption{Population size estimators for simulation study 5.2.}
\centering
\begin{tabular}{rrrrrrrr}
  \hline
Estimator            &Mean      &Var.     &MSE        &Coverage Rate    &Avg. Length           \\\hline
$\hat{N}_{LP}$       &932         &385,736   &390,360        &0.778            &1,859\\
$\hat{N}_1$          &1,111       &589,491    &601,739       &0.759            &10,413 \\
$\hat{N}_3$          &1,207       &1,330,707  &1,373,644     &0.842            &$\infty$ \\
$\hat{N}_5$          &1,072       &508,176    &513,289       &0.674            &2,289 \\
$\hat{N}_{Bayes,\ 1\ wave}$    &640         &38,077    &167,665        &0.467            &642\\
$\hat{N}_{Bayes,\ 2\ waves}$   &924         &10,525     &16,335         &0.833            &365\\
\hline
\label{5_2_Pop_size_ests}
\end{tabular}
\end{table}

\begin{table}[H]
\caption{Bayes data augmentation estimators for simulation study 5.2. Prior 0 used for prior distribution on population size. Estimates are based on one wave.}
\centering
\begin{tabular}{rrrrrrrr}
  \hline
Estimator            &Mean   &Var.                   &MSE         &Coverage Rate  &Avg. Length     \\\hline
$\hat{\lambda}_1$       & 3.592E-01 & 5.415E-03 & 8.457E-03 & 8.196E-01 & 2.430E-01 \\
$\hat{\lambda}_2$       & 2.732E-01 & 5.110E-03 & 5.426E-03 & 8.782E-01 & 2.263E-01 \\
$\hat{\lambda}_3$       & 3.676E-01 & 5.847E-03 & 7.245E-03 & 8.584E-01 & 2.479E-01 \\
$\hat{\beta}_{1,1}$     & 2.434E-02 & 1.657E-04 & 3.166E-04 & 5.956E-01 & 3.213E-02 \\
$\hat{\beta}_{1,2}$     & 2.775E-03 & 3.942E-06 & 7.191E-06 & 6.456E-01 & 5.264E-03 \\
$\hat{\beta}_{1,3}$     & 2.609E-03 & 1.985E-06 & 4.655E-06 & 5.946E-01 & 4.544E-03 \\
$\hat{\beta}_{2,2}$     & 2.543E-02 & 2.945E-04 & 5.746E-04 & 4.502E-01 & 4.213E-02 \\
$\hat{\beta}_{2,3}$     & 3.157E-03 & 5.056E-06 & 9.999E-06 & 5.108E-01 & 5.890E-03 \\
$\hat{\beta}_{3,3}$     & 1.732E-02 & 1.292E-04 & 2.598E-04 & 3.482E-01 & 2.607E-02 \\
\hline
\label{5_2_prior_0_one_wave_results}
\end{tabular}
\end{table}

\begin{table}[H]
\caption{Bayes data augmentation estimators for simulation study 5.2. Prior 0 used for prior distribution on population size. Estimates are based on two waves.}
\centering
\begin{tabular}{rrrrrrrr}
  \hline
Estimator            &Mean   &Var.                   &MSE         &Coverage Rate  &Avg. Length     \\\hline
$\hat{\lambda}_1$        & 3.275E-01 & 1.218E-03 & 1.771E-03 & 9.004E-01 & 1.361E-01 \\
$\hat{\lambda}_2$        & 2.810E-01 & 1.657E-03 & 1.757E-03 & 9.138E-01 & 1.424E-01 \\
$\hat{\lambda}_3$        & 3.915E-01 & 1.721E-03 & 1.904E-03 & 9.298E-01 & 1.577E-01 \\
$\hat{\beta}_{1,1}$      & 1.281E-02 & 3.335E-06 & 3.914E-06 & 9.480E-01 & 6.898E-03 \\
$\hat{\beta}_{1,2}$      & 1.210E-03 & 1.032E-07 & 1.596E-07 & 8.946E-01 & 1.139E-03 \\
$\hat{\beta}_{1,3}$      & 1.161E-03 & 6.004E-08 & 9.486E-08 & 9.178E-01 & 9.815E-04 \\
$\hat{\beta}_{2,2}$      & 1.077E-02 & 6.843E-06 & 1.114E-05 & 8.832E-01 & 8.865E-03 \\
$\hat{\beta}_{2,3}$      & 1.314E-03 & 1.109E-07 & 2.560E-07 & 7.874E-01 & 1.244E-03 \\
$\hat{\beta}_{3,3}$      & 7.349E-03 & 2.200E-06 & 4.324E-06 & 8.496E-01 & 5.532E-03 \\
\hline
\label{5_2_prior_0_two_wave_results}
\end{tabular}
\end{table}

\end{document}